\documentclass{egpubl-no-name}
      
\JournalSubmission

\usepackage[T1]{fontenc}
\usepackage{dfadobe}  

\usepackage{cite}  
\BibtexOrBiblatex
\electronicVersion
\PrintedOrElectronic
\ifpdf \usepackage[pdftex]{graphicx} \pdfcompresslevel=9
\else \usepackage[dvips]{graphicx} \fi

\usepackage{egweblnk} 
  
\graphicspath{{./figs/}{./figs-bert/}}
\usepackage{amsfonts}
\usepackage{soul}
\newcommand {\mm}[1] {\ifmmode{#1}\else{\mbox{\(#1\)}}\fi}

\newcommand{\Rspace}        {\mm{\mathbb{R}}}
\newcommand{\Xspace}        {\mm{\mathbb{X}}}

\newcommand{\Ucal}        {\mm{\mathcal U}}
\newcommand{\Vcal}        {\mm{\mathcal V}}

\newcommand{\myupdate}[1]{{\textcolor{black}{#1}}}

\newcommand{\mapper}{\mm{\mathcal M}}

\newcommand{\para}[1]        {\noindent{\textbf{#1}}}

\newcommand{\etal}{\textit{et al.}}
\newcommand{\ie}{\textit{i.e.}}

\newcommand{\topoact}  {{\mbox{\textbf{TopoAct}}}}

\newcommand{\Nerve}{\mathcal{N}}

\title[TopoAct: Visually Exploring the Shape of Activations]
{TopoAct: Visually Exploring the Shape of Activations\\ in Deep Learning}

\author[A. Rathore, N. Chalapathi, S. Palande, \& B. Wang]
{\parbox{\textwidth}{\centering Archit Rathore$^{1}$, 
		Nithin Chalapathi$^{1}$, Sourabh Palande$^{1}$, Bei Wang$^{1}$\thanks{e-mail: beiwang@sci.utah.edu}
	}
	\\
	{\parbox{\textwidth}{\centering $^1$ School of Computing, Scientific Computing and Imaging (SCI) Institute, University of Utah, USA}
	}
}
\begin{document}
\maketitle

\begin{abstract}
	Deep neural networks such as GoogLeNet, ResNet, and BERT have achieved impressive performance in tasks such as image and text classification.
To understand how such performance is achieved, we probe a trained deep neural network by studying neuron activations, \ie, combinations of neuron firings, at various layers of the network in response to a particular input. 
With a large number of inputs, we aim to obtain a global view of what neurons detect by studying their activations. 
In particular, we develop visualizations that show the shape of the activation space, the organizational principle behind neuron activations, and the relationships of these activations within a layer.
Applying tools from topological data analysis, we present {\topoact}, a visual exploration system to study topological summaries of activation vectors. 
We present exploration scenarios using {\topoact} that provide valuable insights into learned representations of neural networks. 
We expect {\topoact} to give a topological perspective that enriches the current toolbox of neural network analysis, and to provide a basis for network architecture diagnosis and data anomaly detection. 
  
\end{abstract}


\section{Introduction}
\label{sec:introduction}
Deep convolutional neural networks (CNNs) have become ubiquitous in image classification tasks thanks to architectures such as GoogLeNet and ResNet. 
Meanwhile, transformer-based models such as BERT are now the state-of-the-art language model for text classification. 
However, we do not quite understand how these networks achieve their impressive performance. 
One main challenge in deep learning is \emph{interpretability}: How can we make the representations learned by these networks interpretable to humans?

Given a trained deep neural network, we address the interpretability issue by probing neuron activations, \ie, 
the combinations of neurons firings, in response to a particular input image. 
With a large number of input images for a CNN, we can obtain a global view of what the neurons have learned by studying neuron activations in a particular layer. 
We aim to address the following questions: What is the shape of the activation space? What is the organizational principle behind neuron activations? And how are the activations related within a layer and across layers?
We propose to leverage tools from topological data analysis (TDA) to capture global and local patterns of how a trained network reacts to \myupdate{a large number} of input images. In this work:
\begin{itemize}
\item We present {\topoact}, an interactive visual analytics system that uses topological summaries to explore the space of activations in deep learning classifiers for a fixed layer of the network. {\topoact} leverages the mapper construction~\cite{SinghMemoliCarlsson2007} from TDA to capture the overall shape of activation vectors for interactive exploration. 
\item We present exploration scenarios where {\topoact} helps us discover valuable, sometimes surprising, insights into learned representations of image classifiers such as InceptionV1~\cite{SzegedyLiuJia2015} and ResNet~\cite{HeZhangRen2016}. 
\item We observe structures in the topological summaries, specifically branches and loops, that correspond to evolving activation patterns that help us understand how a neural network reacts to a large group of images. In particular, we find a correlation between semantically meaningful distinctions and topological separations among images from different classes.
\item We further demonstrate the generality and utility of {\topoact} by applying it to activation vectors obtained from text classifiers such as the BERT family of models. Via a collaboration with a machine learning expert, we provide concrete use cases in the wild where {\topoact} reveals syntactic and semantic regularities within layers of BERT that help with hypothesis generation in natural language processing (NLP).
\end{itemize}
Finally, we release an open-source, web-based implementation of the exploration interface on Github: {\url{https://github.com/tdavislab/TopoAct/}}; the current system is also available via a public demo link: {\url{https://tdavislab.github.io/TopoAct/}}. 

We expect {\topoact} to benefit the analysis and visualization of neural networks by providing researchers and practitioners the ability to probe black box neural networks from a novel topological perspective. 
\begin{itemize}
\item To the best of our knowledge, {\topoact} is the first tool that focuses on exploring complex topological structures -- branches and loops -- within the space of activation vectors. Its exploratory nature helps to inform the global and local organizational principles of activation vectors across different scales. 
\item {\topoact} detects if and when activations from different classes become separated, via branches in a fixed layer (see \autoref{sec:results}, and in particular, the deer-horse example in \autoref{sec:CIFAR}), which may be used to inform diagnostic or corrective actions such as selective data augmentation for misclassified inputs or network layer modification for increasing the separation between confounding classes (see \autoref{sec:discussion}). 
\end{itemize}

\section{Related Work}
\label{sec:related-work}

We review visual analytics systems for deep learning interpretability as well as various notions of topological summaries. 

\para{Visual analytics systems.}
Visual analytics systems have been used to support model explanation, interpretation, debugging, and improvement for deep learning in recent years; see~\cite{HohmanKahngPienta2018} for a survey. 
Here we focus on approaches based on neuron activations for interpretability in deep learning.  
This line of research attempts to explain the internal operations and the behavior of deep neural networks by visualizing the features learned by hidden units of the network.
Erhan \etal~proposed \emph{activation maximization}~\cite{ErhanDumitruBengio2009}, which uses gradient ascent to find the input image that maximizes the activation of the neuron under investigation. 
It has been used to visualize the hidden layers of a deep belief network~\cite{ErhanDumitruBengio2009} and deep auto-encoders~\cite{Le2013}.
Simonyan \etal~\cite{SimonyanVedaldiZisserman2014} used a similar gradient-based approach to obtain salience maps by projecting neuron activations from the fully connected layers of the CNN back on to the input space.
Building on the idea of activation maximization, Zieler \etal~\cite{ZeilerFergus2014} proposed a deconvolutional network that reconstructs the input of convolutional layers from its output.
Yosinski \etal~\cite{YosinskiCluneNguyen2015} introduced the \emph{DeepVis}  framework that visualizes the live activations produced on each layer of a CNN as it processes images/videos. Their framework also enabled visualizing features in each layer via regularized optimization.

These methods assume that each neuron specializes in learning one specific type of feature.
However, the same neuron can be activated in response to vastly different types of input images.
Reconstructing a single feature visualization, in such cases, leads to an unintelligible mix of color, scales or parts of objects.
To address this issue, Nguyen \etal~\cite{NguyenYosinkiClune2016} proposed \emph{multifaceted feature visualization}, which synthesizes a visualization of each type of input image that activates a neuron.
Another problem with these visualization approaches is the assumption that neurons operate in isolation.
This problem is addressed by the \emph{model inversion} method proposed by Mahendran \etal~\cite{MahendranVedaldi2015,MahendranVedaldi2016}.
Model inversion looks at the representations learned by the fully connected layers of a CNN, and reconstructs the input from these representations.
Kim \etal~introduced the \emph{TCAV} (Testing with Concept Activation Vectors) framework, which uses directional derivatives of activations to quantify the sensitivity of model predictions to an underlying high-level concept~\cite{KimWattenbergGilmer2018}. 
All these techniques can help us understand how a single input or a single class of inputs is ``seen" by the network, but visualizing activations of neurons alone is somewhat limited in explaining the global behavior of the network.
To obtain a global picture of the network, 
Karpathy~\cite{Karpathy2014} used t-SNE to arrange input images that have similar CNN codes (i.e.,~fc7 CNN features) nearby in the embedding. Nguyen \etal~\cite{NguyenYosinkiClune2016} projected the training set images that maximally activate a neuron into a low-dimensional space, also via t-SNE. 
They clustered the images using k-means in the embedded space, and computed a mean image by averaging the images nearest to the cluster centroid. 

Carter \etal~recently proposed the \emph{activation atlas}~\cite{CarterArmstrongSchubert2019}, which combines feature visualization with dimensionality reduction (DR) to visualize averaged activation vectors with respect to millions of input images.
For a fixed layer, the \emph{activation atlas} obtains a high-dimensional activation vector corresponding to each input image. 
These high-dimensional vectors are then projected onto low-dimensional space via UMAP~\cite{McInnesHealyMelville2018,McInnesHealySaul2018} or t-SNE~\cite{MaatenHinton2008}.  
Finally, feature visualization is applied to averaged activation vectors from small patches of the low-dimensional embedding that allow users to intuitively understand how a particular layer reacts to millions of input images. 
Hohman \etal~proposed \emph{SUMMIT}~\cite{HohmanParkRobinson2020}, another framework that summarizes neuron activations of an entire layer of a deep CNN using DR. 
In addition to aggregated activations, \emph{SUMMIT} also computes neuron influences to construct an \emph{attribution graph}, which captures relationships between neurons across layers. 

\emph{Activation atlas} computes average activation vectors in a low-dimensional embedding, which may introduce errors due to neighborhood distortions. 
In comparison, our approach aggregates activation vectors in a different manner.
Using the \emph{mapper} construction, a tool from TDA, we obtain a topological summary of a particular layer by preserving the clusters as well as relationships between the clusters in the original high-dimensional activation space. Our approach preserves more neighborhood structures since the topological summary is obtained within the high-dimensional activation space.
We then study how a particular layer of the neural network reacts to a large number of images through the lens of this topological summary. 

\para{Various notions of topological summaries.}
In TDA, various notions of topological summaries have been proposed to understand and characterize the structure of a scalar function $f: \Xspace \to \Rspace$ defined on some topological space $\Xspace$. 
Some of these, such as merge trees, contour trees~\cite{CarrSnoeyinkAxen2003}, and Reeb graphs~\cite{Reeb1946}, capture the behavior of the (sub)level sets of a function.
Others, including Morse complexes and the Morse-Smale complexes~\cite{EdelsbrunnerHarerZomorodian2003,EdelsbrunnerHarerNatarajan2003}, focus on the behavior of the gradients of a function.
Fewer topological summaries are applicable for a vector-valued function, including Jacobi sets~\cite{EdelsbrunnerHarer2002,BhatiaWangNorgard2015}, Reeb spaces~\cite{EdelsbrunnerHarerPatel2008,MunchWang2016}, and their discrete variant, the mapper construction~\cite{SinghMemoliCarlsson2007}. 
In this paper, we apply the mapper construction to the study of the space of activations to generate topological summaries suitable for interactive visualization. 
The mapper construction introduced by~Singh \etal~\cite{SinghMemoliCarlsson2007} has seen widespread  applications in data science, including cancer research~\cite{NicolauLevineCarlsson2011,MathewsNadeemLevine2019},  sports analytics~\cite{Alagappan2012}, gene expression analysis~\cite{JeitzinerCarriereRougemont2019}, micro-epidemiology~\cite{Knudson2020}, genomic profiling~\cite{Cho2019}, and neuroscience~\cite{GeniesseSpornsPetri2019,SaggarSpornsGonzalez-Castillo2018}, to name a few; see~\cite{PataniaVaccarinoPetri2017} for an overview. 
In visualization, topological approaches such as persistent homology and mapper have recently been applied in graph visualization~\cite{HajijWangRosen2018,HajijWangScheidegger2018,SuhHajijWang2020}.

\section{Comparison with t-SNE and UMAP}
\label{sec:comparison}

Various dimensionality reduction (DR) techniques have been proposed to analyze and visualize high-dimensional point cloud data~\cite{Cayton2008,LiuMaljovecWang2017}.
Among these, t-SNE~\cite{MaatenHinton2008} and UMAP~\cite{McInnesHealyMelville2018} are most relevant to our proposed work as they have been used previously for exploring neuron activations~\cite{Karpathy2014,NguyenYosinkiClune2016,CarterArmstrongSchubert2019}.
In particular, Carter \etal~\cite{CarterArmstrongSchubert2019} employ both t-SNE and UMAP to project high-dimensional activation vectors to low-dimensions for visual exploration. 
In comparison with t-SNE and UMAP, the benefit of {\topoact} is two-fold.
\begin{itemize}
\item {\topoact} provides a global, graph-based representation of the space of activations, which explicitly summarizes the organizational principles (clusters and cluster relations) behind neuron activations. t-SNE and UMAP detect structures visible in the \emph{low-dimensional} embedding, whereas  {\topoact} captures complex topological structures -- loops and branches -- in the original \emph{high-dimensional} space. 
\item Whereas t-SNE and UMAP focus on preserving proximities within  local neighborhoods, {\topoact} explicitly reveals branches and loops that are not necessarily visible via t-SNE/UMAP. These topological structures can be used to guide refined, local structural analysis  (\autoref{sec:results}).   
\end{itemize}
The kNN (k-nearest neighbor) graph constructed by UMAP can be considered as a topological representation of the high-dimensional data~\cite{McInnesHealyMelville2018}. However, it only approximately preserves the \emph{connectivity} among points within local patches of the manifold, and does not capture structures such as loops or branches.
{\topoact} addresses this important challenge by utilizing the mapper construction. 

In addition, several DR techniques have been developed recently~\cite{SilvaMorozovVejdemo-Johansson2009,  WangSummaPascucci2011, YanZhaoRosen2018} that explicitly preserve loops and branches; however, none of them give a global summary of all such structures in a single visualization.
In studying the shape of the space of images, Lee \etal~\cite{LeePedersenMumford2003}, Carlsson \etal~\cite{CarlssonIshkhanovDe-Silva2008}, and Xia~\cite{Xia2016}  studied the global topological structures of patches from natural images; they used different topological tools (such as persistence homology) and focused on different data problems (such as studying the global structure of natural images from a database) in comparison to {\topoact}. 
Finally, a few recent works applied topological techniques in the study of activations. 
Gebhart \etal~\cite{GebhartSchraterHylton2019} computed persistent homology over the activation structure of neural networks. 
In particular, they characterized the topological structure of the neural network architecture when viewed as a graph with edge weights provided by activations. 
Gabella \etal~\cite{GabellaAfamboEbli2019} used both persistent homology and the mapper construction to study the parameter space of neural networks (i.e., weight matrices) during training.  

\section{Technical Background}
\label{sec:background}

We review technical background on the mapper construction and neural network architecture. 
We delay the discussions on activation vectors and feature visualization until~\autoref{sec:methods}.

\para{Mapper graph on point cloud data.} 
We give a high-level description of the framework by Singh \etal~\cite{SinghMemoliCarlsson2007} in a point cloud setting. 
Given a high-dimensional point cloud $\Xspace \subset \Rspace^d$ equipped with a function $f$ on $\Xspace$, $f: \Xspace \to \Rspace$, the mapper construction provides a topological summary of the data for compact representation and exploration. 
It utilizes the topological concept known as the \emph{nerve of a covering}~\cite{Aleksandroff1928}. 

\begin{figure}[!ht]
 \centering
 \includegraphics[width=0.98\columnwidth]{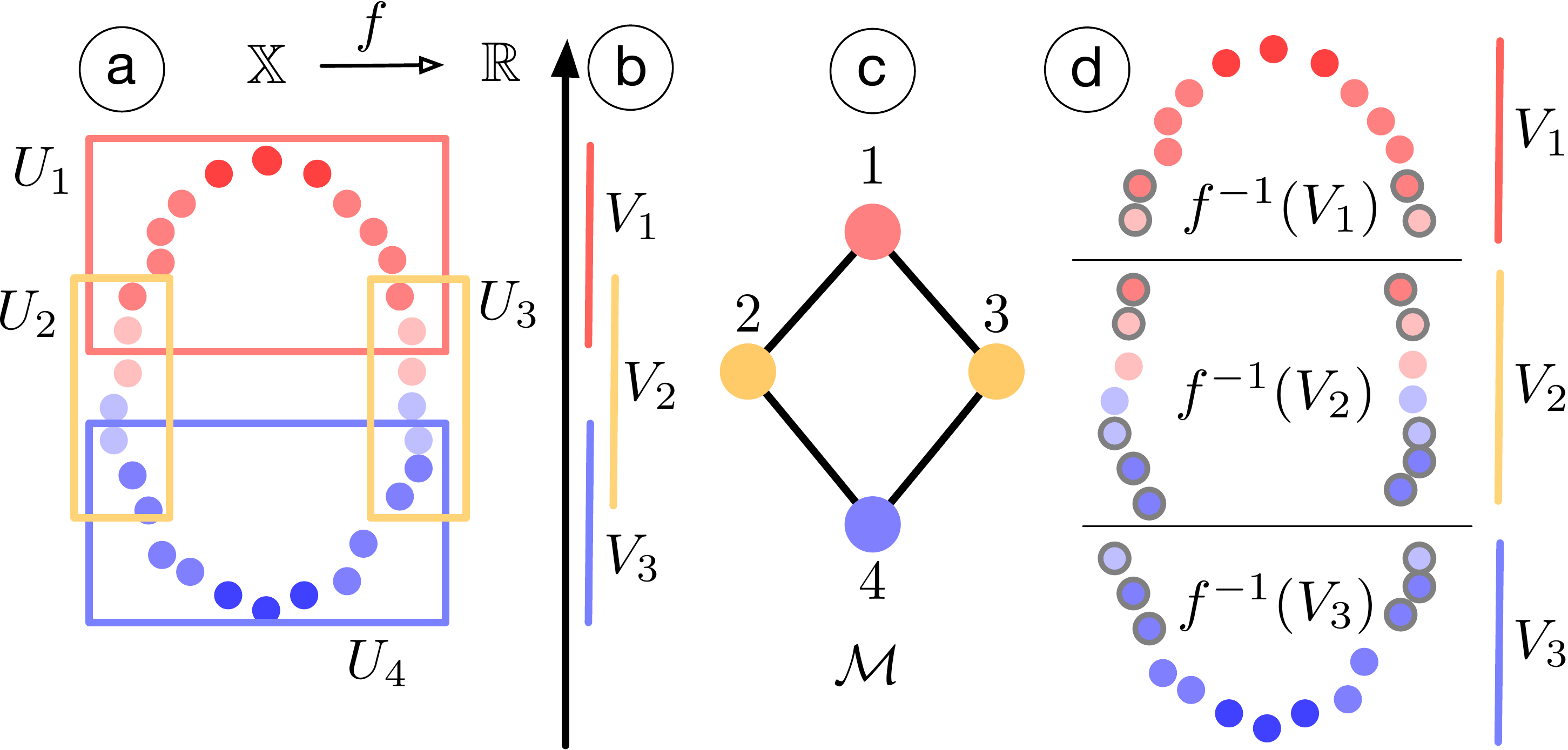}
 \caption{A simple example of a mapper graph on a point cloud.}
 \label{fig:mapper}
\end{figure}

An \emph{open cover} of $\Xspace$ is a collection $\Ucal = \{U_i\}_{i \in I}$ of open sets in $\Rspace^d$ with an index set $I$ such that $\Xspace \subset \bigcup_{i \in I} U_{i}$.
Given a cover $\Ucal$ of $\Xspace$, the $1$-dimensional \emph{nerve} of $\Ucal$, denoted as $\Nerve_1(\Ucal)$, is constructed as follows: A finite set $\{i,j\}\subset I$ (i.e., an edge) belongs to $\Nerve_1(\Ucal)$ if and only if the intersection of $U_i$ and $U_j$ is nonempty; if the set $\{i,j\}$ belongs to $\Nerve_1(\Ucal)$, then any of its subsets (i.e., the point $i$ and the point $j$) is also in $\Nerve_1(\Ucal)$. See~\autoref{fig:mapper} for an example. A cover $\Ucal=\{U_1, U_2, U_3, U_4\}$ that contains open rectangles is given for a 2-dimesional point cloud $\Xspace$ in (a). The $1$-dimensional nerve of $\Ucal$, $\Nerve_1(\Ucal)$, is shown in (c). For instance, there is an edge $\{1,2\}$ that belongs to $\Nerve_1(\Ucal)$ since $U_1 \cap U_2 \neq \emptyset$. 

For the mapper construction, we start with a finite cover $\Vcal = \{V_j\}_{j \in J}$ ($J$ being an index set) of the image $f(\Xspace) \subset \Rspace$ of $f$, such that $f(\Xspace) \subseteq \bigcup_j{V_{j}}$, see~\autoref{fig:mapper}(b).
Since $f$ is a scalar function, $V_i$ is an open interval in $\Rspace$. 
Let $\Ucal$ denote the cover of $\Xspace$ obtained by considering the clusters of points induced by points in $f^{-1}(V_j)$ for each $j$, see (d). 
The $1$-dimensional nerve of $\Ucal$, denoted as $\mapper := \Nerve_1(\Ucal)$, is called the \emph{mapper graph} of $(\Xspace,f)$. 
$\mapper$ is a multiscale representation that serves as a topological summary of of $(\Xspace,f)$, \ie, the point cloud $\Xspace$ equipped with a function $f$. 
Its construction relies on three parameters: the function $f$, the cover $\Vcal$, and the clustering algorithm.   

The function~$f$ plays the role of a \textit{lens}, through which we look at the data, and different lenses provide different insights~\cite{BiasottiGiorgiSpagnuolo2008,SinghMemoliCarlsson2007}.
An interesting open problem for the mapper construction is how to define topological lenses beyond best practices or a rule of thumb~\cite{BiasottiGiorgiSpagnuolo2008,BiasottiMariniMortara2003}. 
In practice, functions such as height, distance from the barycenter, 
surface curvature, integral geodesic distances, and geodesic distances from a source point have all been used as  lenses~\cite{BiasottiGiorgiSpagnuolo2008}. 
In this paper, we use the $L_2$ norm of the activation vectors as the lens, although other options are possible (see the supplementary material for a discussion of $L_2$-norm as a lens function). 

The cover $\Vcal$ of $f(\Xspace)$ consists of a finite number of open intervals as cover elements, $\Vcal=\{V_j\}_{j \in J}$. 
A common strategy is to use uniformly sized overlapping intervals. Let $n$ be the number of intervals and $p$ the amount of overlap between adjacent intervals. Adjusting these parameters increases or decreases the amount of aggregation $\mapper$ provides.

We compute the clustering of the points lying within $f^{-1}(V_i)$ and connect the clusters whenever they have nonempty intersection. A typical algorithm to use is DBSCAN~\cite{EsterKriegelSander1996}, a density-based clustering algorithm; it groups points in high-density regions together and makes points that lie alone in low-density regions outliers. The algorithm requires two input parameters: \emph{minPts} (the number of samples in a neighborhood for a point to be considered as a core point), and \emph{$\epsilon$} (the maximum distance between two samples for one to be considered in the neighborhood of the other).

\autoref{fig:mapper} illustrates a mapper graph construction for a dataset $\Xspace$ sampled from a noisy circle.
The function (lens) used is $f(x) = ||x - p||_2$, where $p$ is the lowest point in the data. 
$\Xspace$ is colored by the value of the function. 
We divide the range of the $f$ into three intervals $\{V_1, V_2, V_3\}$ with a $30\%$ overlap. 
For each interval, we compute the clustering of the points lying within the domain of the lens function restricted to the interval $f^{-1}(V_i)$, and connect the clusters whenever they have a nonempty intersection. 
$\mapper$ is the mapper graph whose nodes are colored by the index set, and it preserves the shape of the point cloud data -- a loop. 

\para{InceptionV1 architecture.}
We give a high-level overview of InceptionV1~\cite{SzegedyLiuJia2015} (GoogLeNet), the neural network architecture employed in this paper. 
However, our framework is not restricted to the specific architecture of a neural network.
InceptionV1 is a CNN that won the ImageNet Large-Scale Visual Recognition Challenge (ILSVRC) for image classification in 2014.  
It was trained on ImageNet ILSVRC~\cite{DengDongSocher2009}.
ImageNet consists of over 15 million labeled high-resolution images with roughly $22K$ classes/categories. ILSVRC takes a subset of ImageNet of around $1K$ images in each of $1K$ classes, for a total of 1 million training images, $50K$ validation images, and $100K$ testing images. 
The highlights of InceptionV1 architecture include the use of $1\times1$ convolutions, inception modules, and global average pooling. 
The $1\times1$ convolution from NIN (networks in networks)~\cite{LinChenYan2014} is used to reduce dimensionality (and computation) prior to expensive convolutions with larger image patches.
A new level of organization is introduced in the form of the \emph{inception module}, which combines different types of convolutions for the same input and stacking all the outputs on top of each other. 
InceptionV1 contains nine \emph{inception modules}, each composed of multiple convolution layers.

The demo version of {\topoact} explores the activations of the last layer of each inception module. 
The module names such as \emph{mixed3a}, \emph{mixed3b} are shortened as \emph{3a}, \emph{3b}, etc.
This choice is well aligned with previous literature on visual exploration of InceptionV1~\cite{OlahMordvintsevSchubert2017,OlahSatyanarayanJohnson2018,HohmanParkRobinson2020,CarterArmstrongSchubert2019}. 

\para{ResNet architecture.}
To demonstrate the generality of topological structures observed across different neural network architectures, we also apply {\topoact} to activation vectors from a ResNet model.  
Residual Network or ResNet~\cite{HeZhangRen2016} was one of first neural network architectures that enabled training extremely deep neural networks with up to $1K$ layers. 
A neural network $N_D$ of depth $D$ is a subnetwork of any network $N_{D+K}$ of depth $D + k, k >0$.
Theoretically, $N_{D+K}$ should be capable of learning any function that $N_D$ can learn by setting the extra $k$ layers to an identity mapping, and thus perform at least as well as the smaller network.
In practice, however, increasing layers beyond a certain depth leads to a sharp degradation in performance (higher training error and lower classification accuracy on the test set) even when normalization schemes are used for both initialization and intermediate representations.
ResNet overcomes this problem by adding ``shortcut'' connections to a layer that adds the output from layer $k$ to the input of layer $k+i$ where $i$ is usually 2.

In our experiments, we have implemented ResNet-18, a residual network with 18 layers following the layer specifications in~\cite{HeZhangRen2016}. With 200 training epochs, it achieves a classification accuracy of 93.24\% on CIFAR-10 and 91.87\% on CIFAR-100 datasets~\cite{Krizhevsky2009}.

\section{Methods}
\label{sec:methods}

We describe data analytic components of {\topoact}.
First, for a chosen layer of a neural network model (such as InceptionV1), we  obtain activation vectors as high-dimensional point clouds for topological data analysis. 
Second, we construct mapper graphs from these point clouds to support interactive exploration.
The nodes in the mapper graphs correspond to clusters of activation vectors in high-dimensional space, and the edges capture relationships between these clusters. 
Third, for each node (cluster) in the mapper graph, we apply feature visualization to individual activation vectors in the cluster and to the averaged activation vector.

\begin{figure}[!ht]
    \centering
    \includegraphics[width=0.90\columnwidth]{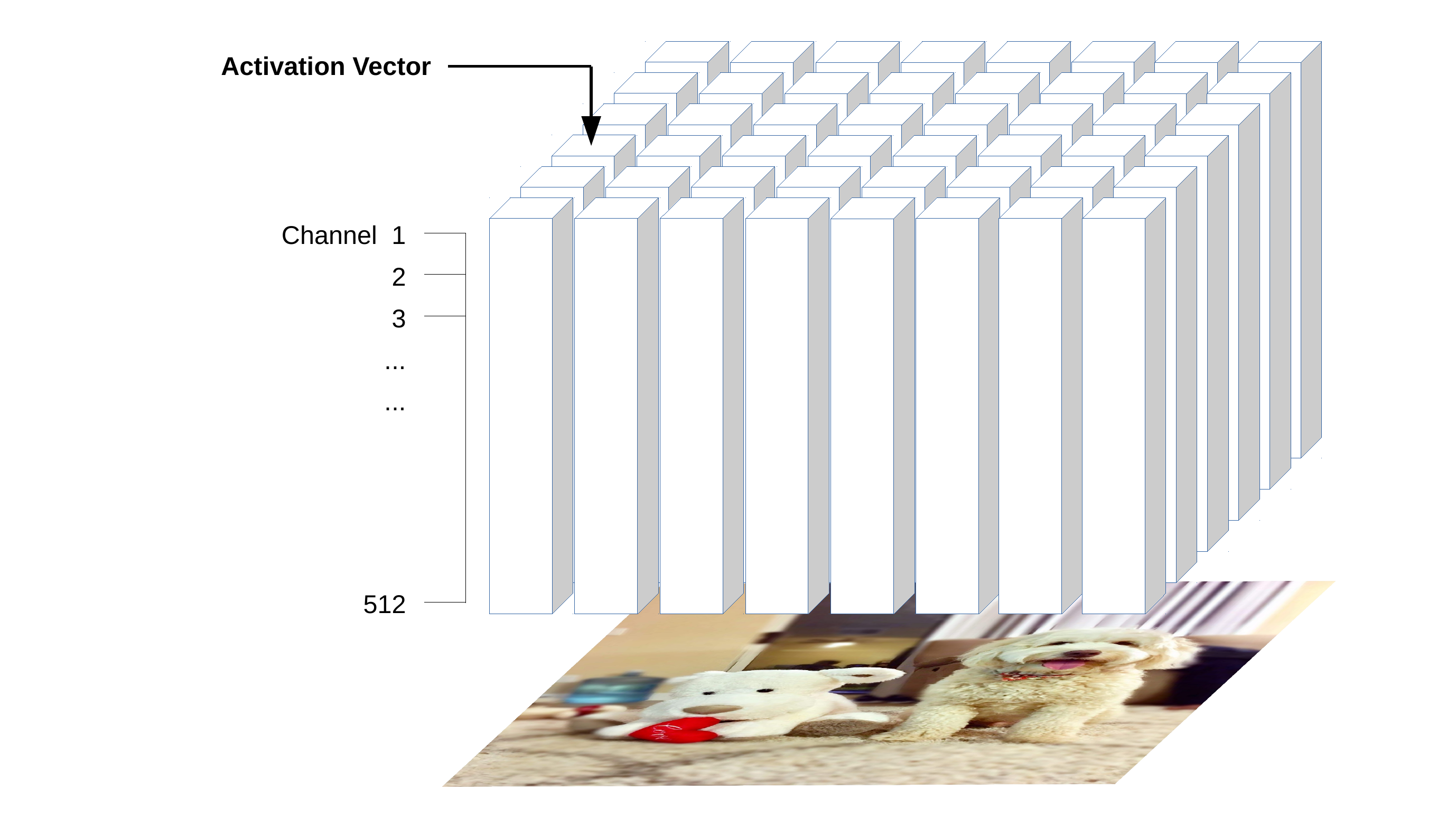}
    \caption{Each activation vector is a vector of spatial activations across all channels.} 
\label{fig:activation-vector}
\end{figure}

\para{Obtaining activation vectors as point clouds.}
The activation of a neuron is a nonlinear transformation of its input.  
To start, we fix a trained model (\ie,~InceptionV1) and a particular layer (~\emph{e.g., 4c}) of interest. 
We feed each input image to the network and collect the activations, \ie, the numerical values of how much each neuron has fired with respect to the input, at a chosen layer.
Since InceptionV1 is a CNN, a single neuron does not produce a single activation for an input image, but instead a collection of activations from a number of overlapping spatial patches of the image. 
When an entire image is passed through the network, a neuron will be evaluated multiple times, once for each patch of the image. 
For example, a neuron within layer \emph{4c} outputs $14 \times 14$ activations per image (for $14 \times14$ patches). 
To simplify the construction, in our setting, we randomly sample a single spatial activation from the $14\times 14$ patches, excluding the edges to prevent boundary effects. 
For $300K$ images, this gives us $300K$ activation vectors for a given layer.
\autoref{fig:activation-vector} illustrates what we mean by an activation vector.
The dimension of an activation vector depends on the number of neurons in the layer. For instance, layers \emph{3a}, \emph{3b}, and \emph{4a} have $256$, $480$, and $512$ neurons, respectively, producing point clouds of corresponding dimensions.

\para{Constructing mapper graphs from activation vectors.}
Given a point cloud of activation vectors, we now compute a mapper graph as its topological summary. 
Each node in the mapper graph represents a cluster of activation vectors, and an edge connects two nodes if their corresponding clusters have a nonempty intersection.  
Our mapper graphs use the $L_2$-norm of the activation vector as the lens function. We set $n = 70$ cover elements with $p= 20\%, 30\%$, and $50\%$ as the amount of overlap. 
We use DBSCAN~\cite{EsterKriegelSander1996} as the clustering algorithm, with minimum points per cluster $\emph{minPts}=5$.
For the $\epsilon$ parameter of the DBSCAN algorithm, which defines core points, we use two variations in our experiments. 
In the first variation, we use a fixed $\epsilon = 1,000$ estimated using the distribution of pairwise distances at the middle layer. 
In the second variation, we set $\epsilon$ adaptively for each layer, employing the procedure proposed in~\cite{EsterKriegelSander1996}.
Specifically, we generate an approximate kNN graph, sort the distances to the $5$-th nearest neighbor, and select an $\epsilon$ based on the location of a ``elbow'' when these distances are plotted~\cite{EsterKriegelSander1996}.
This way, the $\epsilon$ value is more adaptive to 
the activation space of each layer. 
Our \emph{adaptive} $\epsilon$ values are $830$ for \emph{3a}, $1070$ for \emph{3b}, $1750$ for \emph{4a}, $1630$ for \emph{4b}, $1330$ for \emph{4c}, $980$ for \emph{4d}, $775$ for \emph{4e}, $790$ for \emph{5a}, and $260$ for \emph{5b}.

These parameter configurations give rise to six datasets currently deployed in our live demo.
Each dataset contains nine mapper graphs (across nine layers of InceptionV1) constructed by a particular set of parameters associated with the mapper construction. The mapper graphs are named according to these parameters. 
In particular, each name starts with "overlap-$x$" where $x$ is $20$, $30$, or $50$ to denote $20\%$, $30\%$, or $50\%$ overlap, respectively.
The second half of the name consists of "epsilon-$x$" where $x$ is either "fixed" or "adaptive", indicating whether $\epsilon$ was fixed ($=1000$) or set adaptively for each layer.
For example, \emph{overlap-50-epsilon-fixed} is the dataset containing mapper graphs of nine layers generated using $p= 50\%$ with fixed $\epsilon = 1000$. 

\para{Applying feature visualization to activation vectors.}
Activation vectors are high-dimensional abstract vectors. 
We employ \emph{feature visualization} to transform them into a more meaningful semantic representation using techniques proposed by Olah \etal~\cite{OlahMordvintsevSchubert2017, OlahSatyanarayanJohnson2018}.
Whereas the neural network transforms an input image into activation vectors, feature visualization goes in the opposite direction.

Given an activation vector $h_{x,y}$ at a spatial position $(x,y)$, feature visualization synthesizes an idealized image that would have produced $h_{x,y}$ via an iterative optimization process.
Normally, this synthesis is achieved by maximizing the dot product $h_{x,y}\cdot v$ of the vector $h_{x,y}$ with the direction $v$.
However, the vector $v$ that maximizes the dot product can have a large orthogonal component.
To counter this, following~\cite{CarterArmstrongSchubert2019}, the dot product is multiplied with a cosine similarity, putting greater emphasis on the angle between vectors. 
The optimization process, which is similar to back propagation, begins with a random noise image. 
Using gradient descent, this image is iteratively tweaked to maximize the following objective:
${(h_{x,y}\cdot v)^{n+1}}/{(||h_{x,y}|| \,||v||)^n}.$
Subsequently, a \emph{transformation robustness} regularizer~\cite{OlahMordvintsevSchubert2017} is used, which applies small stochastic transformation (jitter, rotate or scale) to the image before applying the optimization step.
Max-pooling can introduce high frequencies in the gradients.
To tackle this problem, the gradient descent is performed in Fourier basis with frequencies scaled to have equal energy, which is equivalent to whitening and de-correlating the data.

Applying feature visualization to all $300K$ activation vectors results in corresponding images that are likely to produce such activations, which we call \emph{activation images}.
Once we obtain a mapper graph, we also apply feature visualization to the averaged activation vector per cluster to obtain an \emph{averaged activation image} for each cluster.
However, feature visualization is not without drawbacks; due to the optimization process and the size of each cluster, it can generate abstract images that remain hard to interpret.
 
\section{Exploring the Shape of Activations from InceptionV1}
\label{sec:results}

\begin{figure*}[tb]
 \centering
  \includegraphics[width=0.9\linewidth]{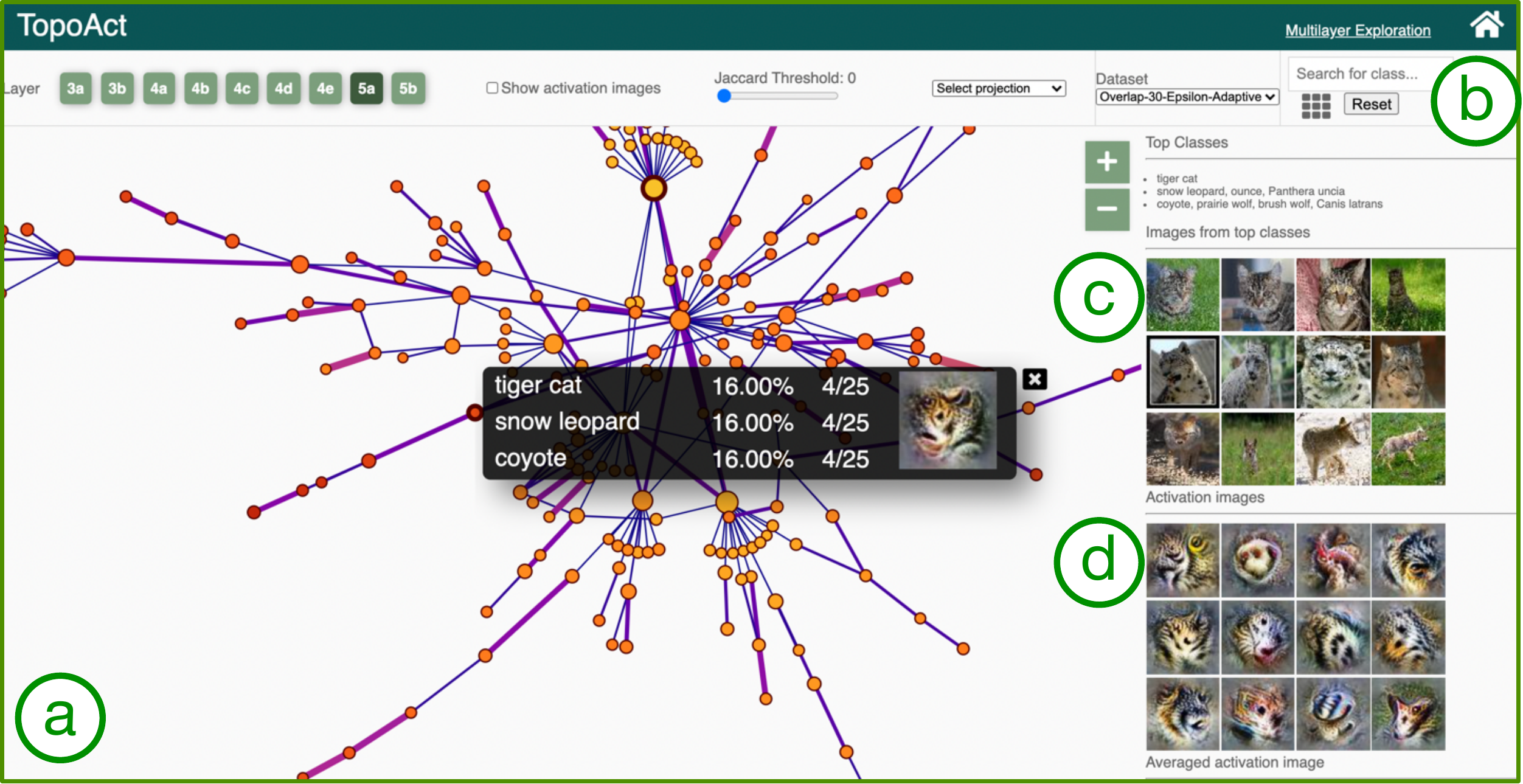}
   \caption{With {\topoact}, users can interactively explore topological summaries of activations in a neural network for a single layer and across multiple layers. Users can investigate activations at a particular layer under the single-layer exploration mode. The mapper graph panel (a) provides a graph-based topological summary of the activation vectors from $300K$ images across $1K$  classes, where each node of the mapper graph represents a cluster of activation vectors, and each edge encodes the relationships between the clusters. The size of a node is proportional to the number of activations, and the color is mapped to the average $L_2$ norm of activations in the cluster. The edge thickness is proportional to the Jaccard index between two nodes.  
  The control panel (b) supports the selection of parameters for the mapper construction and visual encoding.  
   For a chosen cluster in the mapper graph, the data example panel (c)  gives textual description of the top three classes within the cluster together with (up to) five image examples from each top class. The feature visualization panel (d)  applies feature inversion to generate idealized images, called activation images, for individual activation vectors (obtained from data examples) and for an averaged activation vector within a chosen cluster.}
 \label{fig:interface}
\end{figure*}

We present the user interface of {\topoact}, an interactive system used to explore the organizational principles of neuron activations in deep learning image classifiers. 
We use InceptionV1 trained on $1$ million ImageNet images across $1K$ classes. 
We obtain activation vectors of $300K$ images (300 images per class) via the trained model.  
The {\topoact} user interface, \autoref{fig:interface}, contains two exploration modes: single-layer exploration and multilayer exploration. 
We present various exploration scenarios using {\topoact} under the single-layer exploration mode that provide valuable insights into learned representations of InceptionV1. 
See the supplementary materials for a detailed description on the interface,  implementation, and multilayer exploration mode.  

For single-layer exploration, the main takeaway from these scenarios is that {\topoact} captures specific topological structures, in particular, branches and loops, in the space of activations that are hard to detect via classic DR  techniques; such features offer new perspectives on how a neural network ``sees" the input images. 
The topological features identified by {\topoact} can also be used to guide refined, local shape analysis of the space of activations.

\subsection{Discovering Branches from the Space of Activations}

We provide several examples of interesting topological structures  that capture relationships between activations during single-layer exploration. 
Two main types of topological structures unique to our framework, branches and loops, differentiate {\topoact} from prior work (\emph{e.g.},~\cite{HohmanParkRobinson2020,CarterArmstrongSchubert2019}).
Topologically, branches in a graph represent bifurcations, thus separations, among image classes.
Although we observe variations of similar features along a specific branch, different branches may capture distinct, sometimes unrelated, features. 
In order to illustrate the insights gained through user interactions and views between different parts of the system,  figures in this section have average activation images (computed from the average of all activations in a node) overlaid on the nodes of the mapper graph. 

\begin{figure}[!ht]
    \centering
    \includegraphics[width=.99\columnwidth]{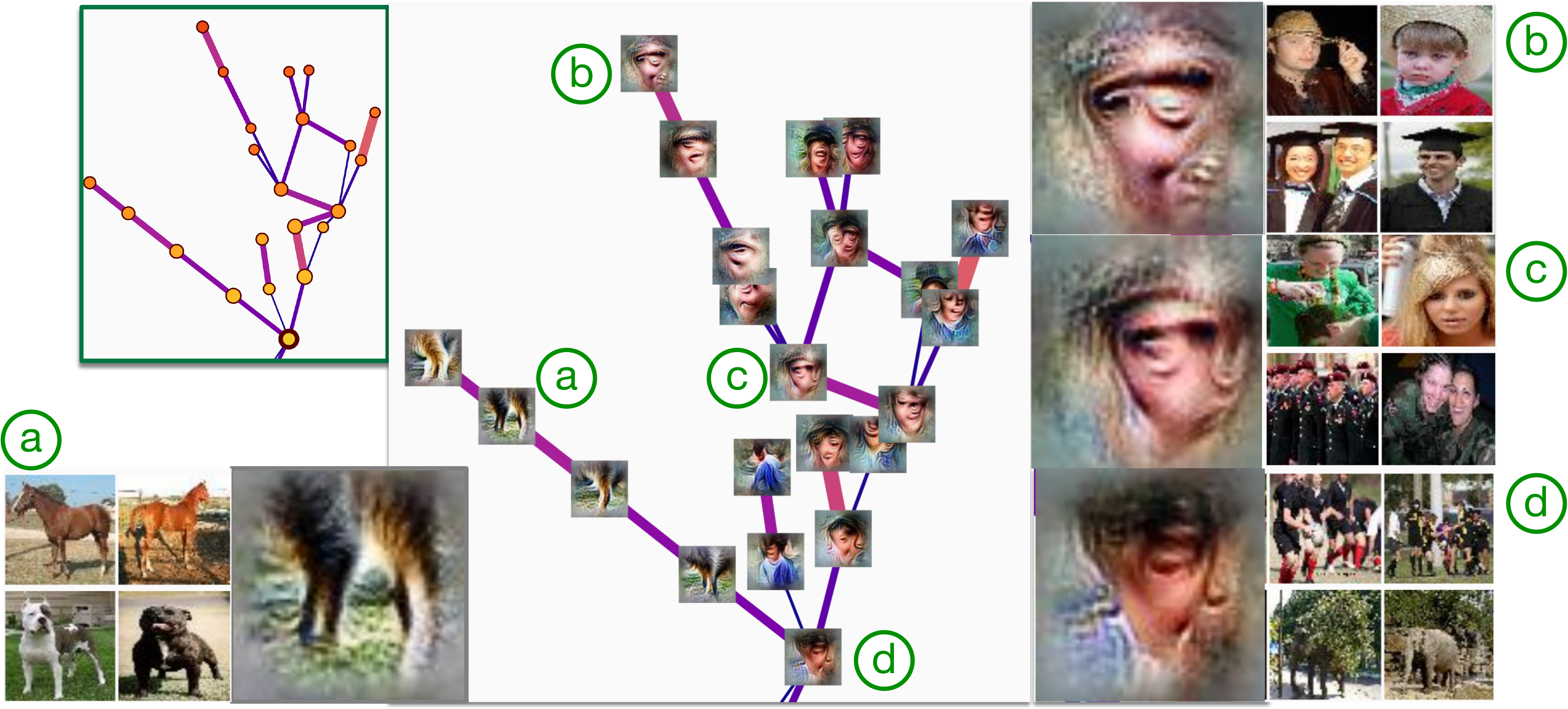}
    \caption{Leg-face bifurcation. Configuration: layer 4c, Euclidean norm, $70$ intervals, $30\%$ overlap, adaptive $\epsilon$ for DBSCAN.} 
\label{fig:branch-leg-face}
   \vspace{-2mm}
\end{figure}

\para{Leg-face bifurcation.}
Our first example of a bifurcation comes from the layer \emph{4c} of the ImageNet dataset (\emph{overlap-30-epsilon-adaptive}). 
\autoref{fig:branch-leg-face} shows two branches emerging from node (d) in the mapper graph; we refer to such a node as the \emph{branching node}.
Node (d) is composed of $381$ activation vectors.
The top three classes within node (d) are \textbf{rugby ball}, \textbf{Indian elephant}, and \textbf{wig}. 
Although this clustering of class labels appears to be random, the mapper graph coupled with averaged activation images reveals interesting insights. 

As illustrated in~\autoref{fig:branch-leg-face}, the left branch appears to capture the leg of an animal.
The top three classes represented in all the clusters within this branch include various breeds of dogs and horses (a).
The right branch appears to capture features that resemble (distorted) human faces.
Although the class names associated with clusters along the right branch may not suggest a relation to human faces, the data examples associated with these clusters reveal that all the top classes in the right branch contain images with humans, most of which include close-ups of faces (b, c).
Returning to the branching node (d), upon closer inspection, we see that it contains images of \textbf{rugby} players and \textbf{elephants} that include both leg and face features, whereas \textbf{wig} images also include human faces. 
Therefore, the activation space bifurcates at the branching node to further differentiate between leg and face features.

\begin{figure}[!ht]
    \centering
    \includegraphics[width=.99\columnwidth]{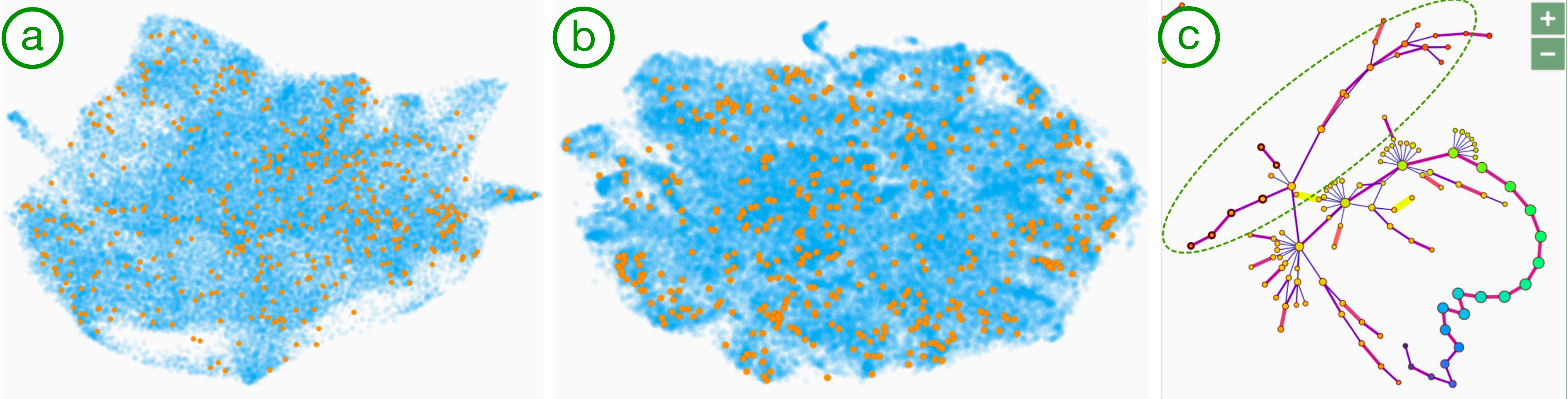}
    \caption{Highlighting activation vectors that belong to the leg-face bifurcation (c) as orange points in the UMAP (a) and t-SNE projection (b).} 
\label{fig:leg-face-projection}
\end{figure}

We further compare {\topoact} against t-SNE and UMAP projections. 
For t-SNE, we use the Multicore-TSNE~\cite{Ulyanov2016} Python library and set the perplexity parameter to be $50$ following the parameter choice used in the \emph{activation atlas}~\cite{CarterArmstrongSchubert2019}. The UMAP projection is performed using its official Python implementation~\cite{McInnesHealySaul2018} with $20$ nearest neighbors and a minimum distance of $0.01$.
As illustrated in \autoref{fig:leg-face-projection}, we select nodes that are involved in the leg-face bifurcation and highlight their corresponding activation vectors in the t-SNE and UMAP projections. 
In particular, neither t-SNE nor UMAP reveals a bifurcation as the activation vectors in the projection are scattered over the entire projection. 

\begin{figure}[!ht]
    \centering
    \includegraphics[width=.99\columnwidth]{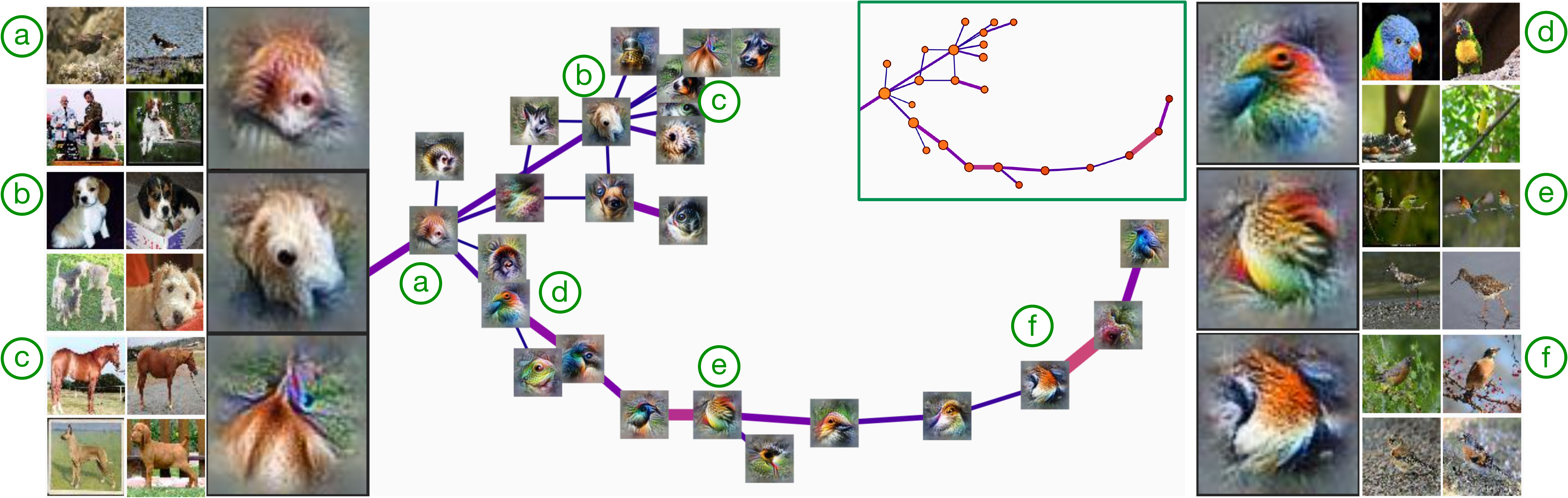}
    \caption{Bird-mammal bifurcation. Configuration: layer 5a, Euclidean norm, $70$ intervals, $30\%$ overlap, fixed $\epsilon$ for DBSCAN.} 
\label{fig:bird-mammal}
\vspace{-4mm}
\end{figure}

\para{Bird-mammal bifurcation.}
Our second example of a branch comes from the layer \emph{5a} of the ImageNet dataset (\emph{overlap-30-epsilon-fixed}). 
\autoref{fig:bird-mammal} shows two branches emerging from the branching node (a), which is composed of $398$ activation vectors. 
It contains images of both birds and dogs such as \textbf{oystercatcher} and \textbf{Brittany spaniel}, and the averaged activation image of the branching node appears to be a combination of the left profile of bird faces and right profile of the dog-like faces.
Upon further investigation, the bottom branch that contains nodes (d), (e), and (f) focuses on the features of bird faces: profile views composed of the left eye and beak, with variations of color and textures as we move along the branch.
The clusters in this branch include mainly bird species such as \textbf{bee eater}, \textbf{robin}, and \textbf{lorikeet}.
The variations in the captured features and corresponding data samples can be seen in nodes (d), (e), and (f).
The clusters in the top branch, on the other hand, appear to capture features of mammalian faces: eyes and snouts, with variations in color and texture.
This branch primarily consists of images from classes of mammals, including various dog breeds, \textbf{wolves}, and \textbf{foxes}.  

\begin{figure}[!ht]
    \centering
    \includegraphics[width=.99\columnwidth]{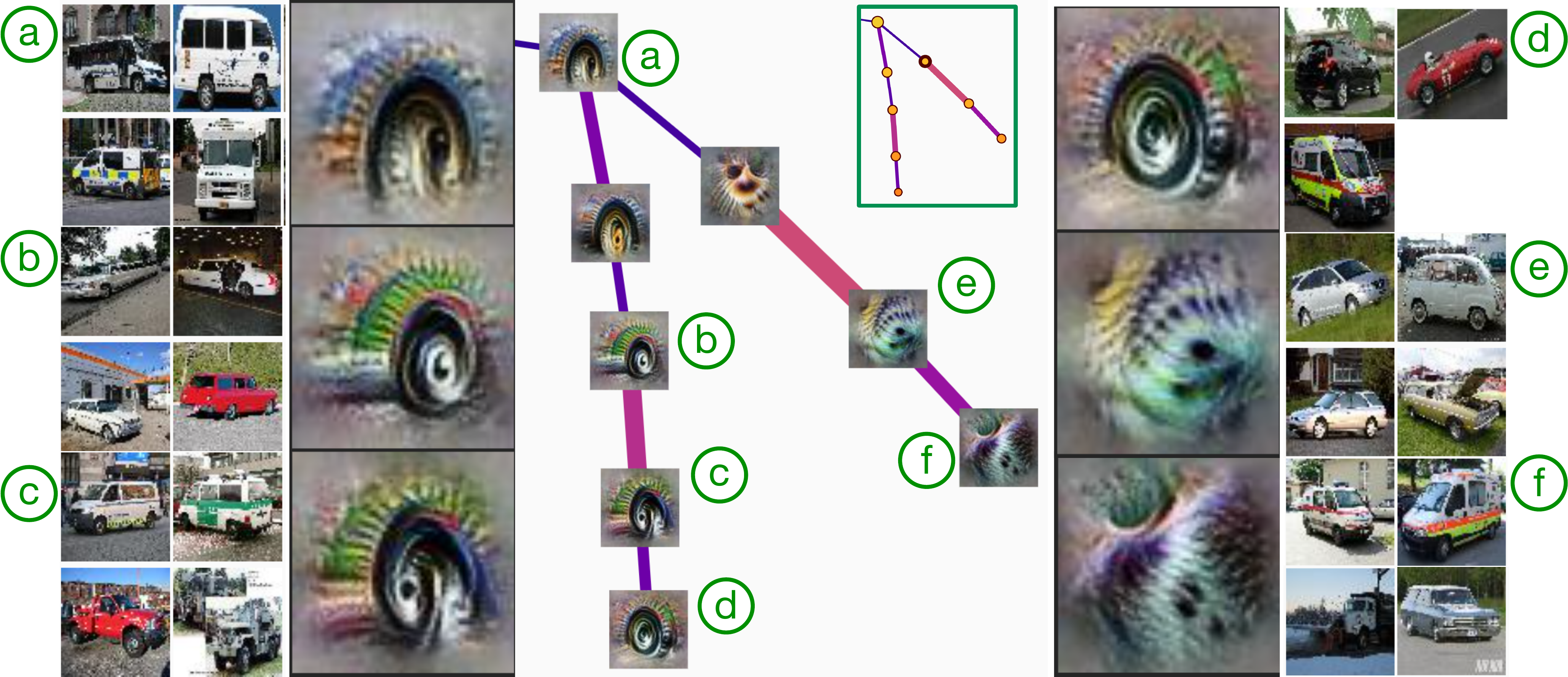}
    \caption{Wheel-tread bifurcation. Configuration: layer 4c, Euclidean norm, 70 intervals, 30 \% overlap, adaptive $\epsilon$ for DBSAN.}
\label{fig:wheel-tread}
\end{figure}

\para{Wheel-tread bifurcation.}
Our third example comes from the layer \emph{4c} of the ImageNet dataset (\emph{overlap-30-epsilon-adaptive}).
As illustrated in~\autoref{fig:wheel-tread}, the branching node (a) is a small cluster of size $136$.
All the clusters in this example contain images of various types of automobiles, for example \textbf{minibus}, \textbf{police van}, \textbf{fire engine}, \textbf{limousine}, etc.
The branching node (a) appears to capture what looks like the wheel of a vehicle - a dark round shape with tread-like pattern.
The two branches appear to focus on one of these two features.
Whereas the left branch focuses on the dark round swirling patterns of automobile wheels (b, c, d), the right branch appears to focus more on the tread-like patterns and textures (e, f).

\subsection{Exploring Loops from the Space of Activations}
Branches capture bifurcations in the types of features across different objects, but some loops seem to capture different aspects of the same underlying object.

\begin{figure}[!ht]
    \centering
    \includegraphics[width=.99\columnwidth]{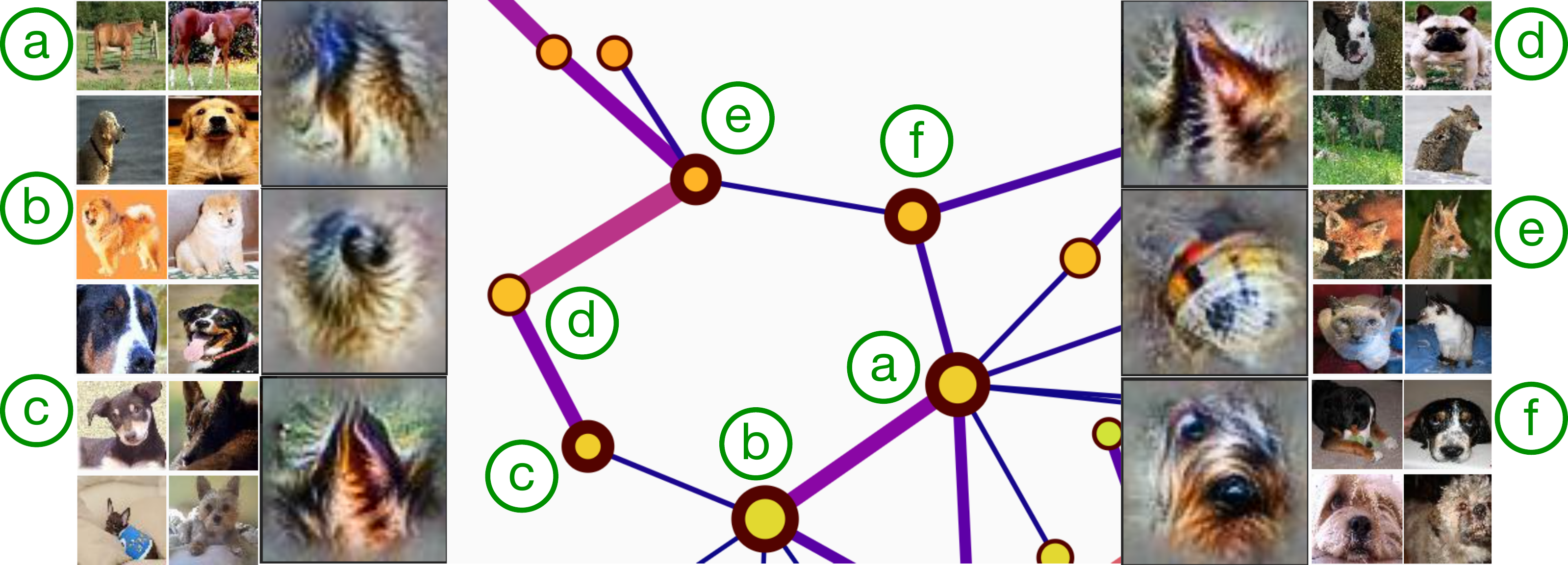}   
    \caption{Fur-nose-ear-eye loop. Configuration: layer 4d, $70$ intervals, $30\%$ overlap, fixed $\epsilon$ for DBSCAN.}
\label{fig:fur-nose-loop}
\end{figure}

\para{Fur-nose-ear-eye loop of mammals.}
Our first example comes from layer \emph{4d} of the ImageNet dataset (\emph{overlap-30-epsilon-fixed}). 
\autoref{fig:fur-nose-loop} shows a loop created by six clusters.
The top classes in all six clusters include various dog breeds and a variety of foxes.
All these clusters seem to capture different features (\ie,~body parts) related to these animals.
Based on feature visualization, the leftmost cluster appears to capture the color patterns and the texture of the fur from the body (a).
Going clockwise, the next cluster also captures the color and texture of the fur but from a different body part, possibly the fur and hair surrounding the nose, suggested by the dark spot and the swirling pattern (b).
The next two clusters (c, d) appear to capture animal ears.
The averaged activation image captured by the cluster (e) is not as clearly attributable to a specific part of an animal's body.
As can be observed in (e), this cluster consists of images from a larger variety of animals, from \textbf{foxes} to \textbf{Siamese cats} and \textbf{hogs}.
As a result, the corresponding averaged activation image is a mixture of various colors and slightly different textures.
The last cluster (f) appears to capture the eyes and noses of the animals.
We can observe in (f) that the cluster contains front and side views of dog heads.

\begin{figure}[!ht]
    \centering
    \includegraphics[width=.99\columnwidth]{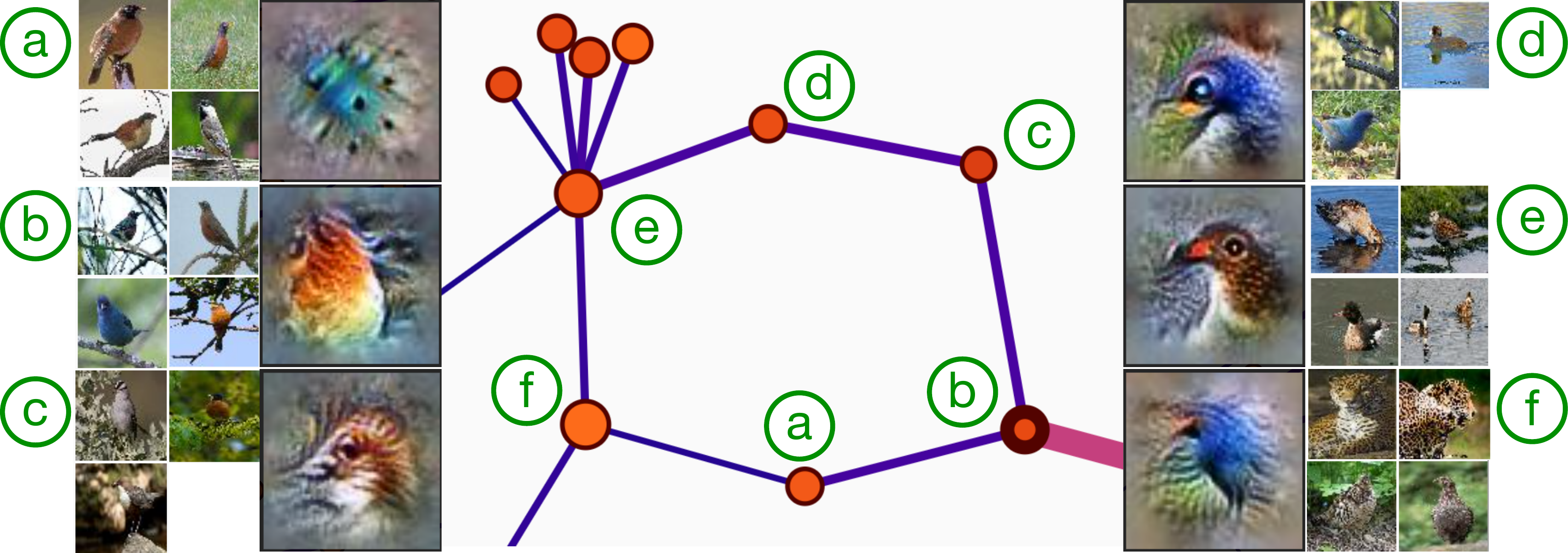}
    \caption{Face-body-leg loop of birds. Configuration: layer 5a, $70$ intervals, $30\%$ overlap, adaptive $\epsilon$ for DBSCAN.} 
\label{fig:face-body-loop}
\end{figure}

\para{Face-body-leg loop of birds.}
Our next example originates from layer \emph{5a} of the ImageNet dataset (\emph{overlap-30-epsilon-adaptive}). 
\autoref{fig:face-body-loop} shows six clusters creating the loop.
The top three classes of all the clusters in the loop consist of bird species, and similar to the previous example, the averaged activation images show us different features (body parts) of the birds captured by these clusters.
Clusters (c, d, e) on the top of the loop appear to capture the left profile views of the bird faces with the left eye and the beak identifiable in the averaged activation images.
These clusters are, in fact, composed of images of birds.
The variation in the color of birds (from red to blue, and to brown) is reflected in the corresponding activation images (b, c, d, e).
Clusters (a, b, f) on the bottom of the loop appear to capture the body and legs along with a feathered texture, although not as clearly as the other three clusters.
As can be seen, cluster (f) also includes images of \textbf{leopard} and \textbf{jaguar} mixed with images of birds (\textbf{partridge} and \textbf{ruffed grouse}) for the representation of texture.

\subsection{Studying Global Views of Activation Spaces}
We now explore the global view of an activation space using the single-layer exploration mode. 
Instead of focusing on a single type of topological structure such as loops or branches, we investigate the distribution of topological structures. 
\begin{figure}[!ht]
    \centering
    \includegraphics[width=.98\columnwidth]{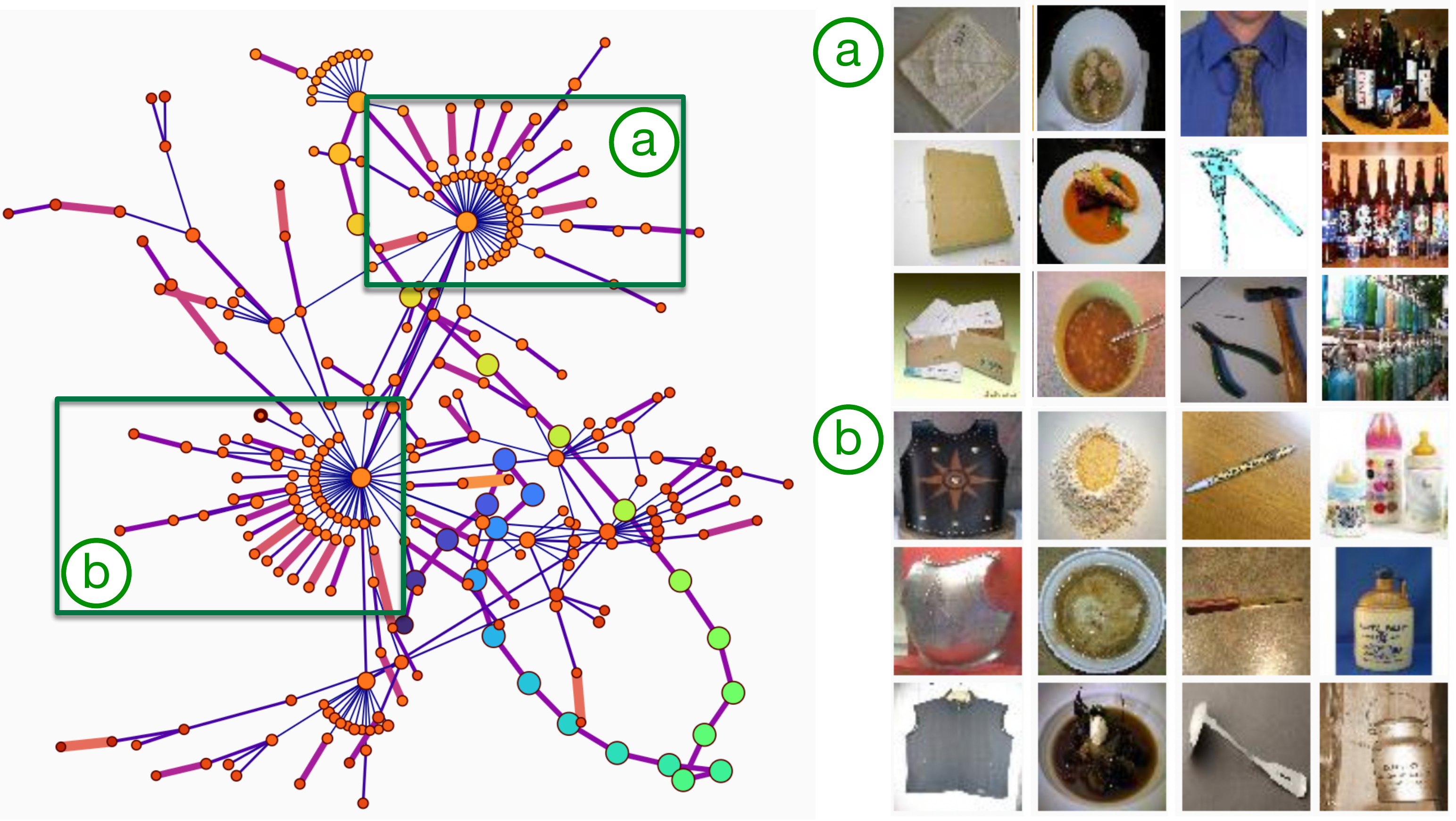}
    \vspace{-2mm}
    \caption{A global view of a mapper graph for a fixed layer. Configuration: layer 5b, $70$ intervals, $30\%$ overlap, adaptive $\epsilon$ for DBSCAN.} 
\label{fig:global-5b}
  \vspace{-4mm}
\end{figure}
As illustrated in \autoref{fig:global-5b}, we investigate the distribution of branches within the largest connected component of a mapper graph, at layer \emph{5b}  of the ImageNet dataset (\emph{overlap-30-epsilon-adaptive}).  
We pay special attention to branching nodes with high degrees (a, b). 
These branching nodes, in some sense, serve as ``anchors'' or ``hubs'' of the underlying space of activations.
We make a few interesting, though speculative, observations.  
For each of the two branching nodes in (a) and (b), a mixture of geometric and texture-based images contributes to the representation of the node. 
Nodes immediately adjacent to the branching node (a), \ie, those that form branches that merge at node (a), contain geometric objects that are square-shaped (\textbf{envelop}, \textbf{bath towel}), circle-shaped (\textbf{bowl}, \textbf{pasta}), pointy-shaped (\textbf{tie}, \textbf{hammer}), and bottle-shaped (\textbf{beer}). 
Nodes immediately adjacent to the branching node (b) have other objects that serve similar purposes, including square-shaped (\textbf{vest}, \textbf{cuirass}), circle-shaped (\textbf{dough}, \textbf{mashed potato}), pointy-shaped (\textbf{ladle}, \textbf{ball point}), and bottle-shaped (\textbf{milk cans}, \textbf{whiskey jug}). 
However, (a) and (b) seem to draw these geometric shapes from (almost completely) different classes of images, which may indicate a level of self-similarity within the space of activations that requires further investigation.

\subsection{Refined Analysis of Topological Structures}
\label{sec:pca}

We can utilize interesting topological structures identified by {\topoact} -- branches and loops -- to obtain topologically meaningful subsets of the activation vectors for further analysis.
We present some examples of {\topoact}-guided principal component analysis (PCA) with the following procedure. 
We first identify all nodes that form a branch or a loop within a mapper graph. 
We then extract activation vectors (as high-dimensional points) that map to these nodes.
Next, we apply PCA to these points and project them to a 2-dimensional plane. 

\begin{figure}[!ht]
    \centering
    \includegraphics[width=.99\columnwidth]{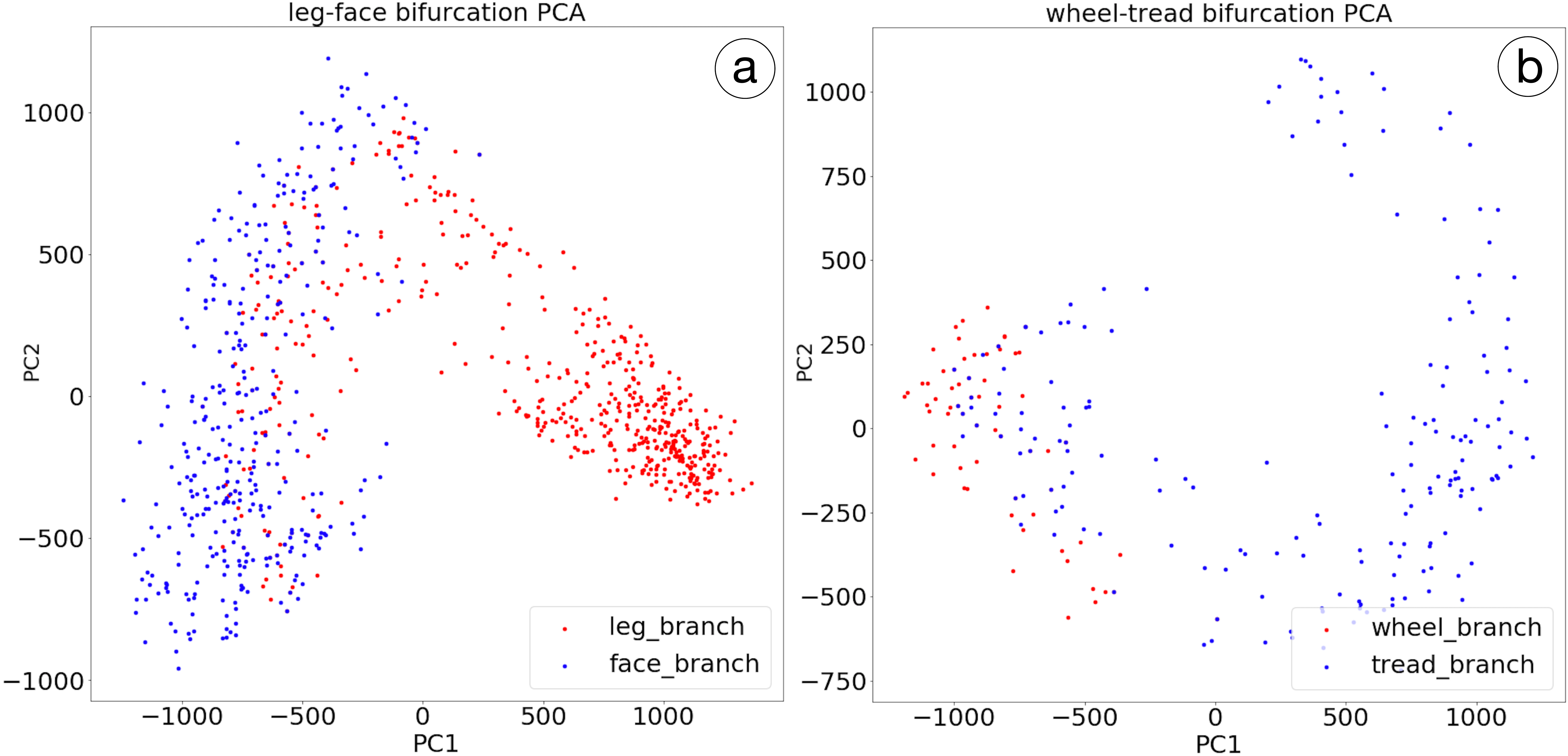}
    \caption{PCA of the activation vectors that belong to (a) the leg-face bifurcation and (b) the wheel-tread bifurcation.} 
\label{fig:pca-leg-face}
\vspace{-4mm}
\end{figure}

Consider the leg-face bifurcation from \autoref{fig:branch-leg-face}.~\autoref{fig:pca-leg-face}(a) shows the PCA projection of all points that participate in the bifurcation. 
The red points belong to the activation vectors from the ``face'' branch and the blue points belong to the ``leg'' branch.
Similarly, for the wheel-tread bifurcation from \autoref{fig:wheel-tread}, ~\autoref{fig:pca-leg-face}(b) illustrates the PCA projection of its associated points. 
For both examples, we could easily observe that points from the two branches lie along two distinct directions. 

\begin{figure}[!ht]
    \centering
    \includegraphics[width=.98\columnwidth]{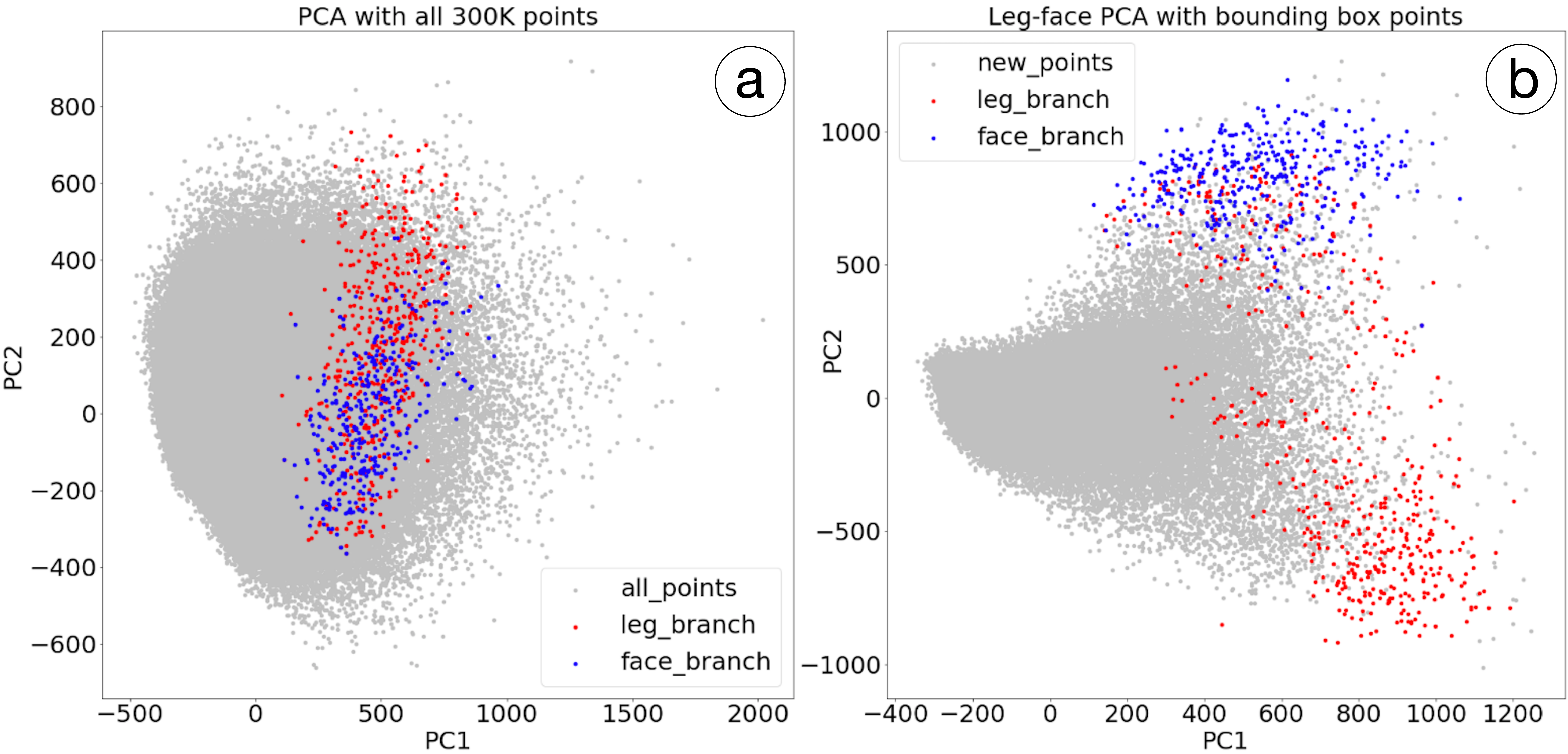}
    \caption{PCA applied to the entire set of activation vectors from layer \emph{4c} (a), and to the activation vectors in the neighborhood of leg-face bifurcation (b). We do not see any branching or clustering structure in (a), while (b) reveals the leg-face bifurcation with respect to its neighboring points.} 
\label{fig:mixed4c-pca}
\end{figure}

For comparative purposes, we apply PCA to all 300K points from layer \emph{4c} and highlight those from the leg-face bifurcation. 
As shown in~\autoref{fig:mixed4c-pca}(a), there are no observable branching or clustering structure within this global projection. 
To verify that the leg-face bifurcation is not spurious, we construct a minimal bounding box and identify around 86K neighboring points in the space of activations. 
We apply PCA to these 86K points and observe that points from the leg-face bifurcation form clusters that are separable from their neighboring points; see ~\autoref{fig:mixed4c-pca}(b), which confirms that the leg-face bifurcation detected by {\topoact} is not spurious.

This example demonstrates the advantage of using {\topoact} in combination with classic DR techniques (such as PCA, t-SNE, and UMAP) to perform refined shape analysis of the space of activations.
Although the total number of activation vectors in our dataset is large, the number of significant branches and loops is relatively small, which leads us to hypothesize that a layer is particularly well-trained to identify certain directions in the activation space.
As shown here, {\topoact} can help us identify these directions.

\section{Applying TopoAct to ResNet Trained on CIFAR}
\label{sec:CIFAR}

To demonstrate the generality of our framework, we provide additional experiments using ResNet trained on the CIFAR-10 and CIFAR-100 datasets~\cite{Krizhevsky2009}. 
Both datasets consist of the same set of $60K$ color images of dimension  $32 \times 32$, with $50K$ training images and $10K$ test images. 
CIFAR-10 has $10$ images classes with $6K$ images per class, and CIFAR-100 has $100$ image classes with $600$ images per class.
The class labels in CIFAR-10 are coarser, such as \textbf{automobiles} and \textbf{mammals}; whereas classes in CIFAR-100 are finer, such as \textbf{bicycle}, \textbf{bus}, \textbf{beaver}, and \textbf{hamster}. 
We demonstrate that the insights provided by {\topoact} are not specific to a particular dataset or a particular network architecture. 
We give a few exploration scenarios involving branches by applying {\topoact} to ResNet-18 trained on the CIFAR-10 dataset; such examples are similar to those described in~\autoref{sec:results}. 
We encourage readers to explore further with our open-source online demo. 

\begin{figure}[!h]
     \centering
     \includegraphics[width=.99\columnwidth]{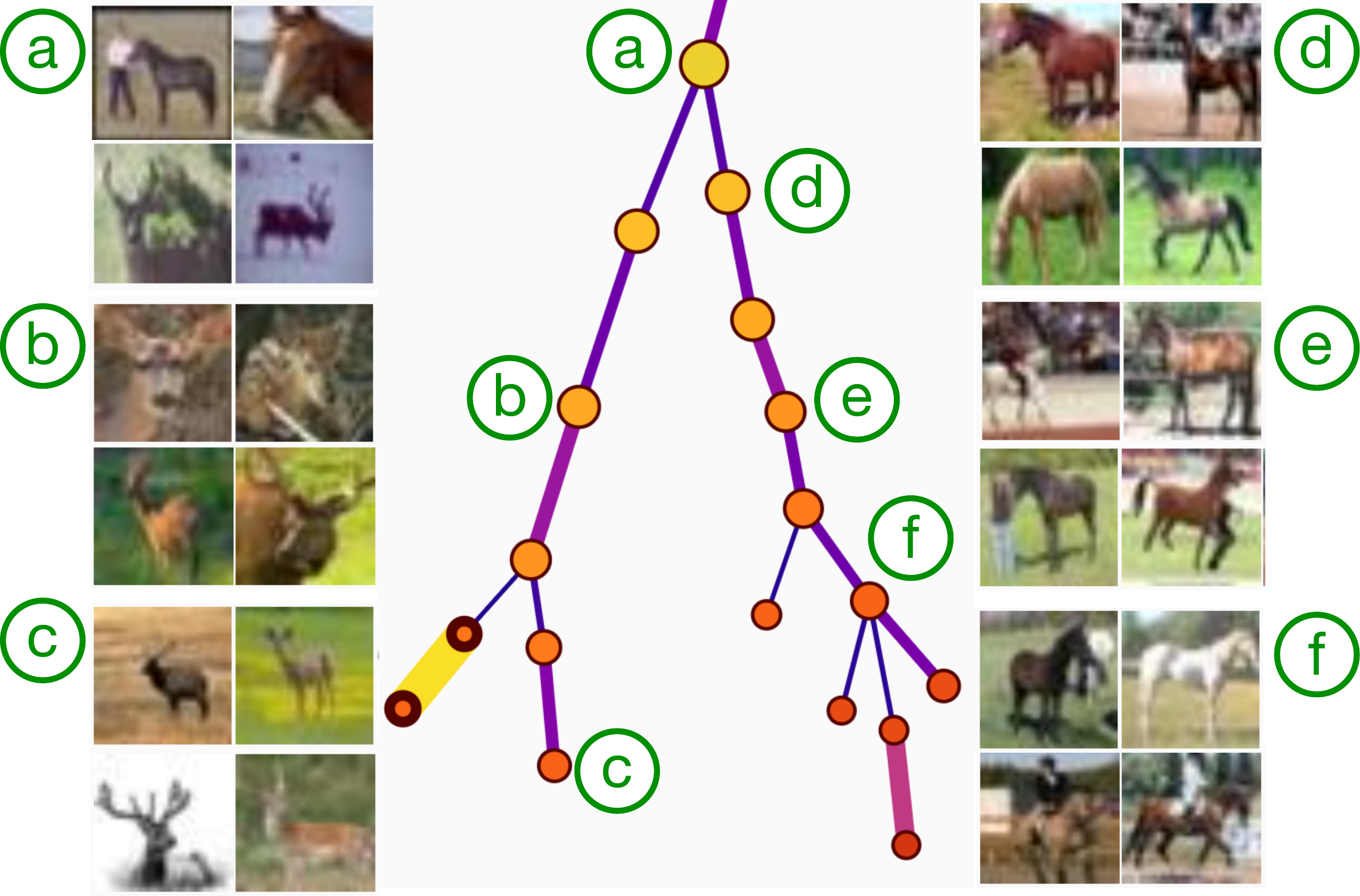}
     \vspace{-2mm}
     \caption{Horse-deer bifurcation. Configuration: layer {4.1.bn2}, $40$ intervals, $20\%$ overlap.} 
       \vspace{-2mm}
 \label{fig:horse-deer}
 \end{figure}

\para{Horse-deer bifurcation.}
Our first example is a horse-deer bifurcation from the last layer \emph{4.1.bn2} of the CIFAR-10  dataset, as illustrated in \autoref{fig:horse-deer}. 
The left branch that contains nodes (b) and (c) corresponds to images of deer, whereas the right branch with nodes (d), (e) and (f) corresponds to images of horses. 
The branching node (a) contains images of both horses and deer. 
In addition, none of the earlier layers show such a clear bifurcation between the horse and deer classes. {\topoact} reveals the layer at which the network first begins to differentiate between these two classes.
Such insights would make {\topoact} a useful diagnostic tool for deep learning researchers (see~\autoref{sec:discussion}).

\begin{figure}[!h]
 \vspace{-2mm}
     \centering
     \includegraphics[width=.99\columnwidth]{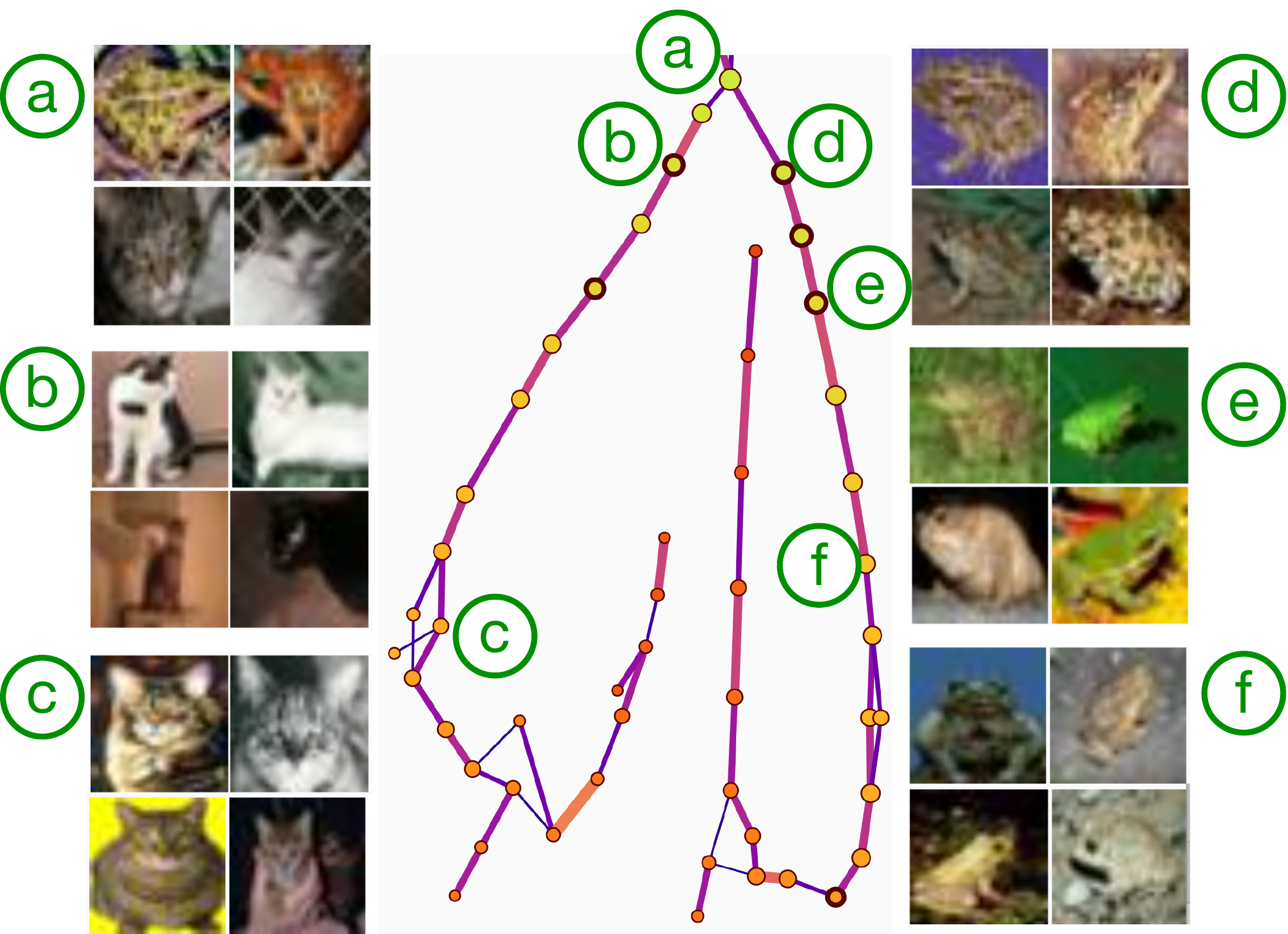}
     \vspace{-2mm}
     \caption{Frog-cat bifurcation. Configuration: layer {4.1.bn2}, $100$ intervals, $40\%$ overlap.} 
 \label{fig:frog-cat}
   \vspace{-2mm}
 \end{figure}

\para{Frog-cat bifurcation.}
Similarly, our second example is a frog-cat bifurcation from the last layer \emph{4.1.bn2} of the CIFAR-10  dataset, as illustrated in \autoref{fig:frog-cat}. 
Here, the branching node (a) contains images of frogs and cats. 
It then bifurcates into a left branch (with nodes (b) and (c)) that contains only images of cats, and a right branch (with nodes (d), (e), and (f)) that contains only images of frogs. 
Even though these are very different types of animals (mammals vs. amphibians), they share similar postures. 

\section{Applying {\topoact} to BERT Neural Network for NLP}
\label{sec:bert}

To demonstrate the utility of {\topoact} with concrete use cases in the wild, we have collaborated with a machine learning (ML) expert in Natural Language Processing (NLP), Dr. Vivek Srikumar from the University of Utah. 
We apply {\topoact} to activation vectors obtained from the BERT (Bidirectional Encoder Representations from Transformers) family of models, which is the default representation of text for a variety of NLP tasks. 

Although BERT and similar models are undoubtedly helpful in improving predictive accuracy, it is not immediately clear why it helps. This problem arises because BERT embeddings (\ie,~activation vectors) consist of a collection of high-dimensional vectors for every sentence, and this high-dimensional space is not easy to explore. 
The highlight of our collaboration is that {\topoact} presents an opportunity in this context by revealing the various syntactic and semantic regularities that each layer of BERT captures, some of which are surprising but interpretable for the ML expert.  

In addition, {\topoact} shows potential directions for improving these representations, for instance, targeting a specific downstream NLP task. 
For example, in addition to some clear patterns of concepts captured via topological branches, we also see clusters and relationships between them that appear noisy, yet stable. 
The existence of these suggests that there may be opportunities for developing topologically aware regularization techniques for training BERT-like models by imposing additional constraints such that the mapper graphs resulting from {\topoact} should exhibit certain structural properties, which is left for future work.

\subsection{BERT and Activation Datasets}
BERT and other transformers-based language models have recently found widespread application as the go-to methods for many tasks in NLP such as sentiment analysis, sentence classification, and domain-specific language modeling \cite{BeltagyLo2019, LeeYoon2020}. Jawahar \etal \cite{JawaharSagot2019} explored and established that contextual embeddings from BERT do indeed encode syntactic structures in earlier layers and compositional structures in later layers. 

For our experiments, we use the ``\emph{bert-base-uncased}'' trained network from the \textit{Huggingface's transformers} library \cite{WolfDebut2019} with 12 layers, each with 768 neurons.
We collect activation vectors from BERT on the training set of the Georgetown University Multilayer (GUM) corpus \cite{Zeldes2017}. The data contains 4780 sentences and $81,857$ tokens.
Tokens are individual words, numbers, or punctuation marks that form sentences. 
Among the $80K$ tokens, about $11K$ are punctuations. 

We collect the activations by passing each sentence through the trained BERT model, collecting per-token activations, and applying {\topoact} to the activations for all 12 layers.  
That is, for each of the 12 layers of the BERT neural network, we compute mapper graphs for point clouds of the token's activations in 768 dimensions across various parameter settings, similar to our earlier examples involving InceptionV1 and ResNet-18.  
 
Our experiments applying {\topoact} to BERT activations confirm and expand upon the earlier results in NLP~\cite{BeltagyLo2019, LeeYoon2020, JawaharSagot2019}. 
In the following sections, we present some use cases of structures found in the BERT activations through our tool that highlight both local and global structures in both syntactic and semantic regimes. 

\subsection{Pronoun Differentiation}
For layer 12 (the last layer) of BERT, we notice a branching  structure that highlights the differentiation among pronouns. 
Recall that a pronoun is ``a word that can function by itself as a noun phrase and that refers either to the participants in the discourse (\emph{e.g.}, \textbf{I}, \textbf{you}) or to someone or something mentioned elsewhere in the discourse (\emph{e.g.}, \textbf{she}, \textbf{it}, \textbf{this})'', according to Oxford English Dictionary.

\begin{figure}[!ht]
\centering{
\includegraphics[width=.98\linewidth]{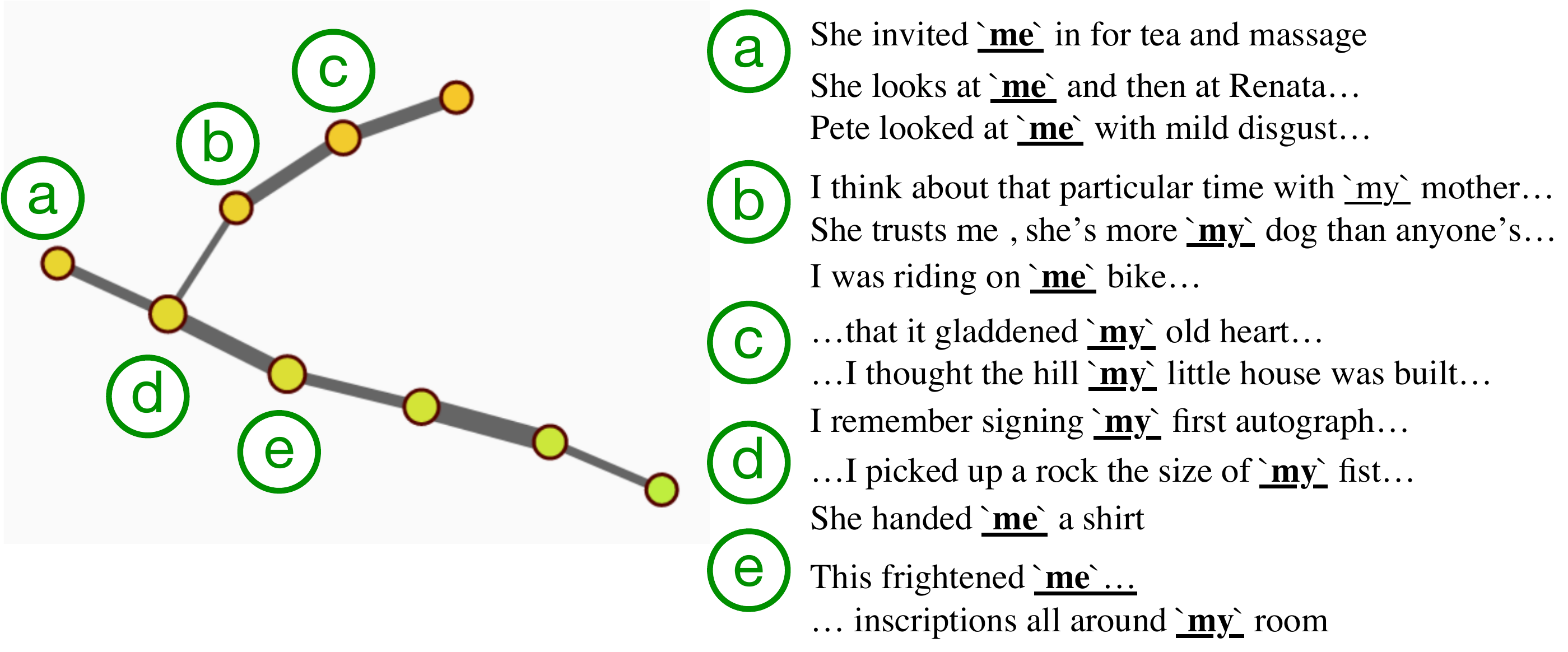}
}
\caption{Pronoun differentiation. Configuration: layer 12, Euclidean norm, 80 intervals, 30\% overlap.}
\label{fig:me-my-pronouns}
\end{figure}

As illustrated in~\autoref{fig:me-my-pronouns}, the chain of nodes with (d) and (e) represents a mixture of the first person singular personal pronoun (i.e, \textbf{me}) and the first person singular personal possessive form (\ie, \textbf{my}). The chain continues along (b) and (c) with the possessive form (\textbf{my}), but the personal pronoun splits into its own branch (a) (\textbf{me}). 
Node (a) consists of sentences which employ the word \textbf{me} in phrases such as \emph{``She invited \textbf{me} in for tea''},\emph{``She looks at \textbf{me} and then at Renata''}, \emph{``He soon came back and gave \textbf{me} the good news''}, etc. 
Node (b) consists of sentences that employ the word \textbf{my} in phrases such as \emph{``since I was in \textbf{my} twenties''}, \emph{``I wished \textbf{my} da would come home''}. 
In particular, it also contains a sentence \emph{``I was riding on \textbf{me} bike and I thought I'd swallowed an insect''}; here the word \textbf{me} is used instead of \textbf{my} likely due to a local dialect.    
This indicates that the differentiation between the pronouns goes beyond just the word-level and instead captures semantic differences as well. 

This structure is interesting to our ML expert for a couple of reasons. First, we have a cluster of only the first person forms (\textbf{me} and \textbf{my}), and not the second (\textbf{you}) and third person forms (\textbf{he, she, they}). Second, although the pronoun (specifically, the object pronoun \textbf{me}) and the possessive form (\textbf{my}) are related in one sense (i.e., first person singular), they are distinct both in terms of their meanings and grammatical roles. 
The branching structure highlights this similarity and their divergence.

\subsection{Contextual Differentiation}
\begin{figure}[!ht]
\centering{
\includegraphics[width=0.95\linewidth]{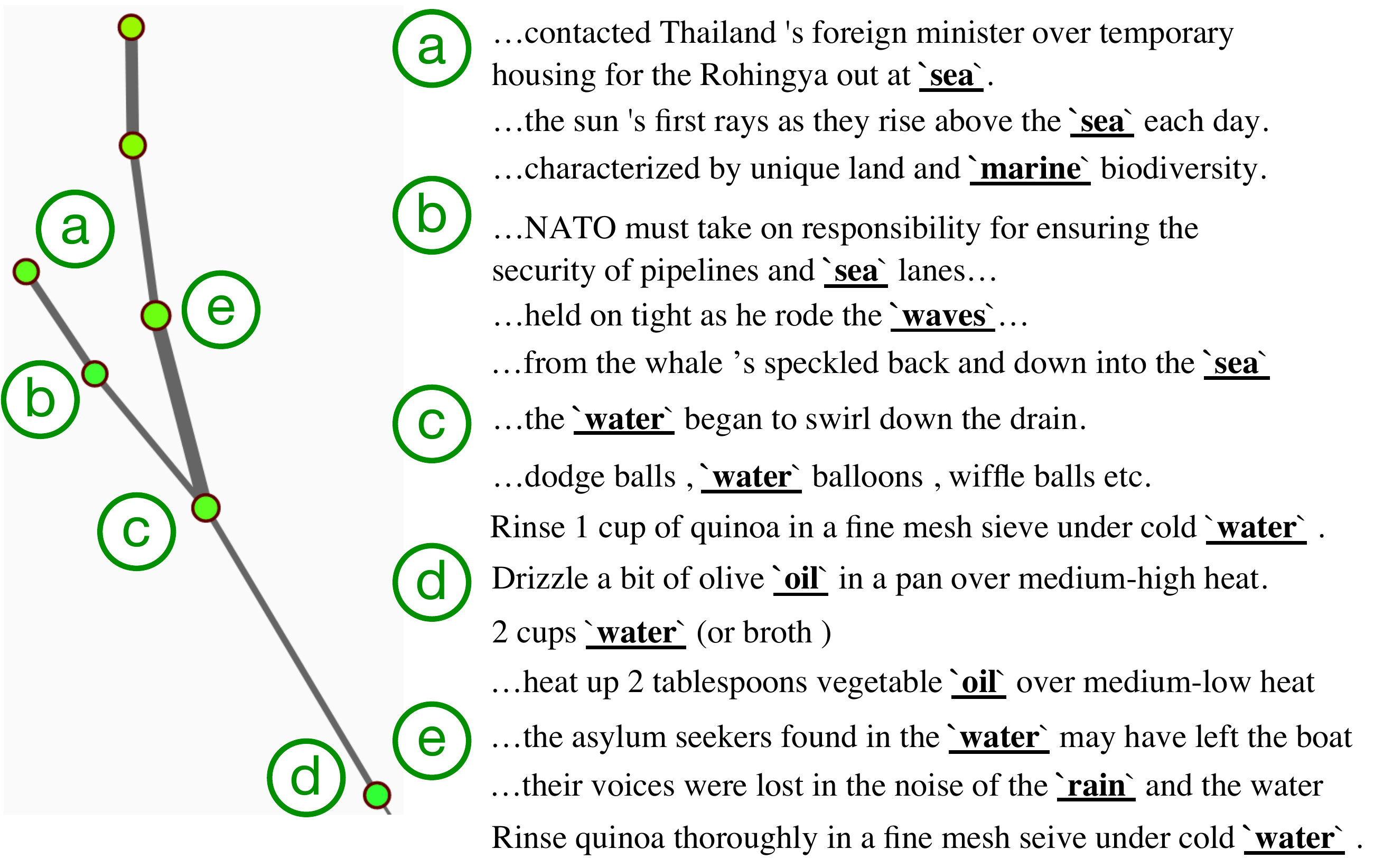}
}
\caption{Contextual differentiation. Configuration: layer 9, Euclidean norm, 80 intervals, 30\% overlap.}
\label{fig:water-sea}	
\end{figure}

We give two examples involving how branches from {\topoact} capture contextual differentiations. 
As shown in~\autoref{fig:water-sea}, the structure of nodes about water  highlights the different roles that water may play.
Nodes (a) and (b) reflect oceanic usages (\emph{e.g.},~\textbf{sea}, \textbf{marine}, \textbf{waves}).
Node (d) reflects culinary usages (\emph{``2 cups \textbf{water} (or broth)''}) and starts to connect with other liquids used in cooking (e.g., \emph{``olive \textbf{oil} to taste}''). 
And other labeled nodes (\emph{e.g.},~node (e)) reflect meteorological usages (\emph{``the noise of the \textbf{rain}''}), culinary usages (\emph{``rinse...under cold \textbf{water}''}), and oceanic usages.  

\begin{figure}[!ht]
\centering{
	\includegraphics[width=0.98\linewidth]{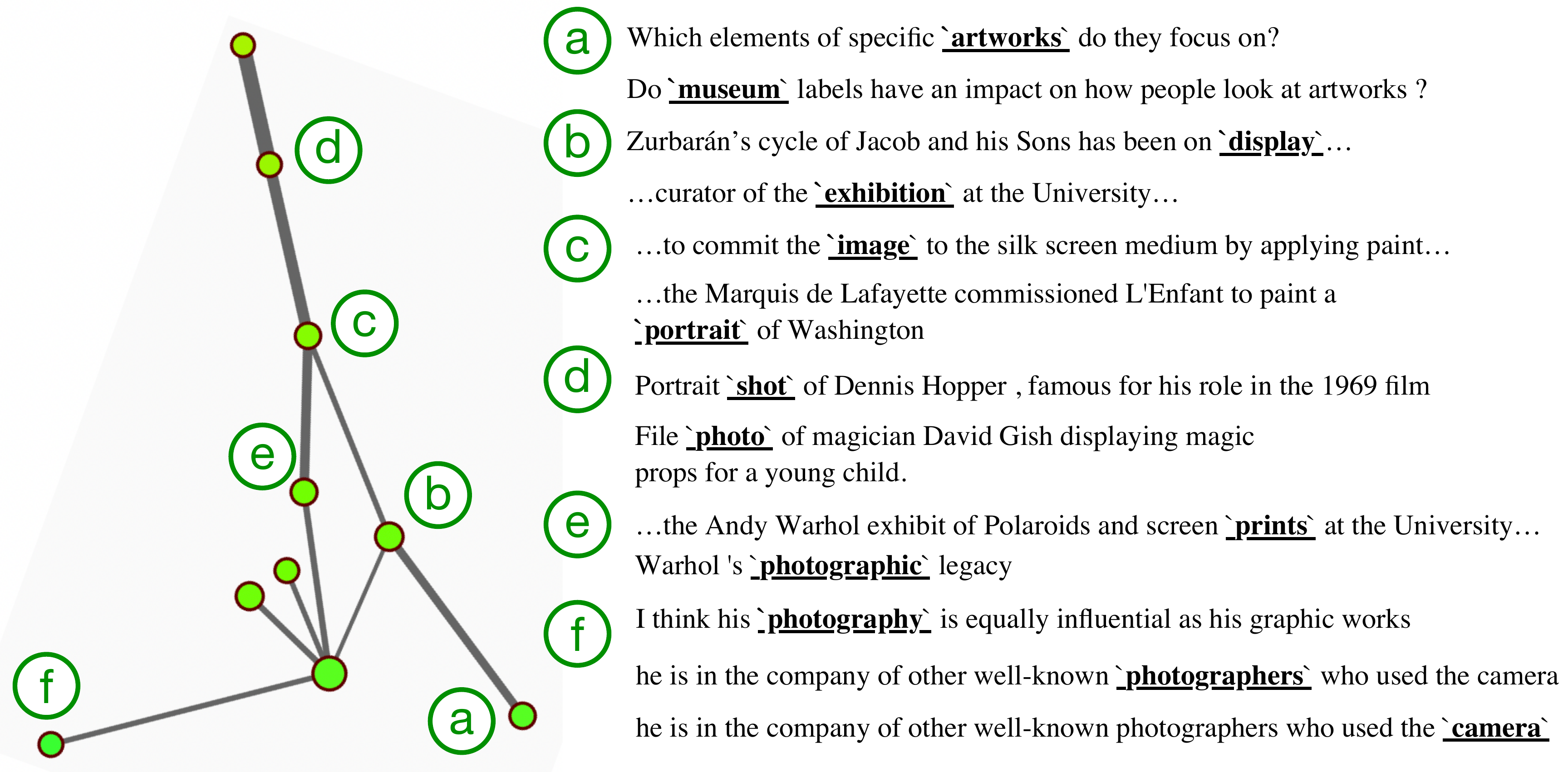}
}
\caption{Contextual differentiation. Configuration: layer 9, Euclidean distance, 80 intervals, 30\% overlap, Jaccard $=0.01$.}
\label{fig:art}		
\end{figure}

We see similar fine-grained distinctions between \textbf{photographs}, \textbf{photography}, and \textbf{art} in~\autoref{fig:art}. 
The structure starts with \textbf{artworks}, \textbf{museum}, and \textbf{art} in node (a), moves on to \textbf{display}, \textbf{exhibition}, and \textbf{exhibits} in node (b), which then gets further refined into  \textbf{portrait} and \textbf{painting} in node (c), as well as \textbf{shot}  and \textbf{photo} in node (d).

The fact that BERT (without any fine-tuning for any task) captures these differences is surprising for our ML expert. 
Such an insight could point toward an explanation for how transformer models like BERT and its many variants seem to capture world knowledge and even common-sense knowledge~\cite{BosselutRashkin2019}. 

\subsection{Local and Global Syntax}

\begin{figure}[!ht]
\centering{
\includegraphics[width=0.99\linewidth]{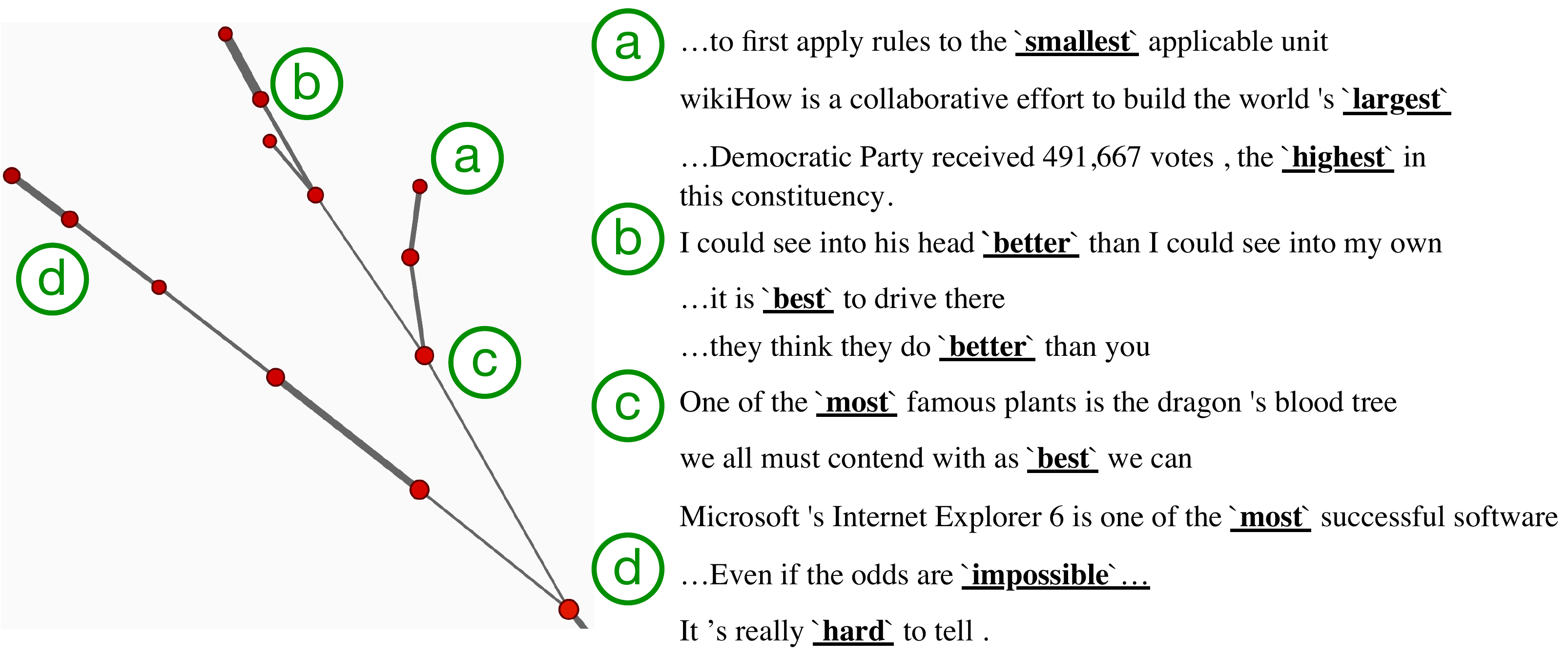}
}
\caption{Local syntax at an earlier layer. Configuration: layer 3, cosine distance, 80 intervals, 30\% overlap.}
\label{fig:local-syntax}	
\end{figure}

It is known that the lower layers of BERT characterize more local syntax and non-contextual lexical semantics, whereas the later layers capture global sentential structure and semantics~\cite{TenneyDas2019}. 
We see this by comparing the structures we see in layer 3 in (\autoref{fig:local-syntax}) with the ones in layer 9 (\autoref{fig:me-my-pronouns}, \autoref{fig:water-sea}, \autoref{fig:art}). 

In lower layers, we observe groupings of words with similar parts of speech that diverge into chains that contain only words with similar meaning, as shown in~\autoref{fig:local-syntax}. 
For example, a generic adjective node (c) bifurcates into a set of nodes that describe size (e.g., \textbf{bigger}, \textbf{largest}, \textbf{highest} in node (a)) and a set of nodes that describe goodness (e.g., \textbf{best} and \textbf{better} in node (b)).

The other examples from later layers -- such as layer 9 (\autoref{fig:global-syntax}) -- show that the mapper graph encodes complex relational abstractions that go beyond simply the dictionary meaning of the word. 

\subsection{After-When Separation}
\begin{figure}[!ht]
\centering{
\includegraphics[width=0.99\linewidth]{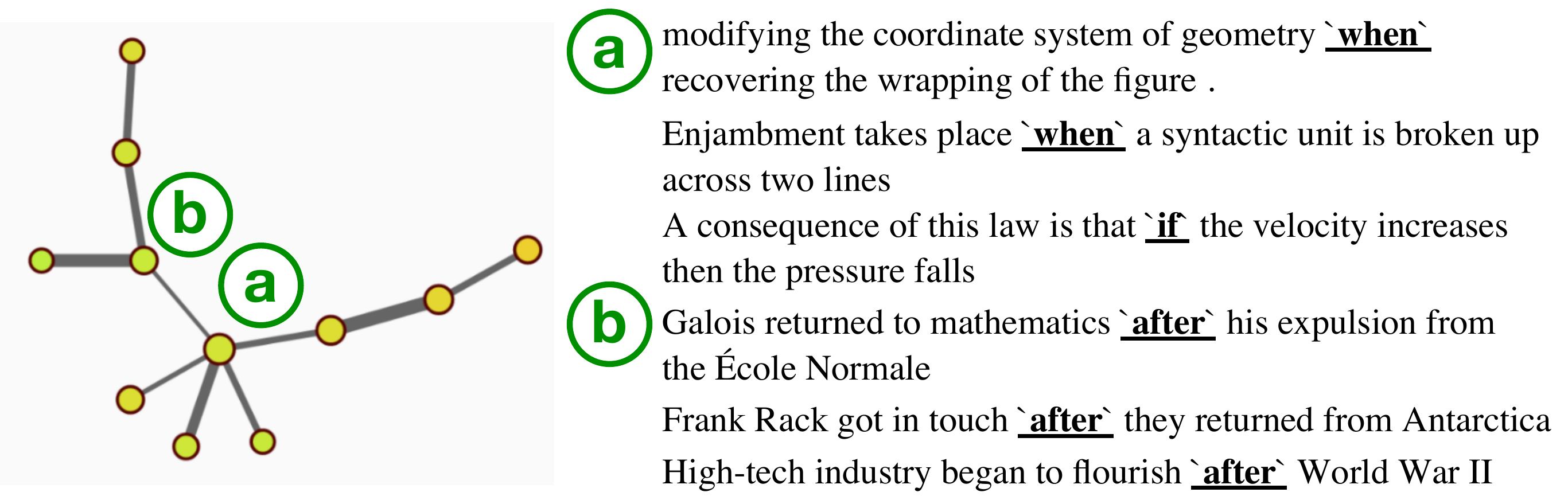}
}
\caption{After-When separation. Configuration: layer 9, Euclidean distance, 80 intervals, 40\% overlap.}
\label{fig:global-syntax}
\end{figure}

The branching structure in~\autoref{fig:global-syntax} represents a surprising and difficult to characterize dichotomy in the usage of words like “after” and “when”. Both these words are used to convey temporal meaning, but, the latter is also used to introduce discourse relationships such as explanations or sometimes even causations. The fact that these two usages are distinctly represented in the BERT embeddings shows that BERT does indeed characterize this subtle difference and, as a result, could serve as a basis for developing future parsers for discourse or rhetorical structure of text.

\subsection{Temporal and Locative Prepositions}

\begin{figure}[!ht]
\centering{
\includegraphics[width=0.99\linewidth]{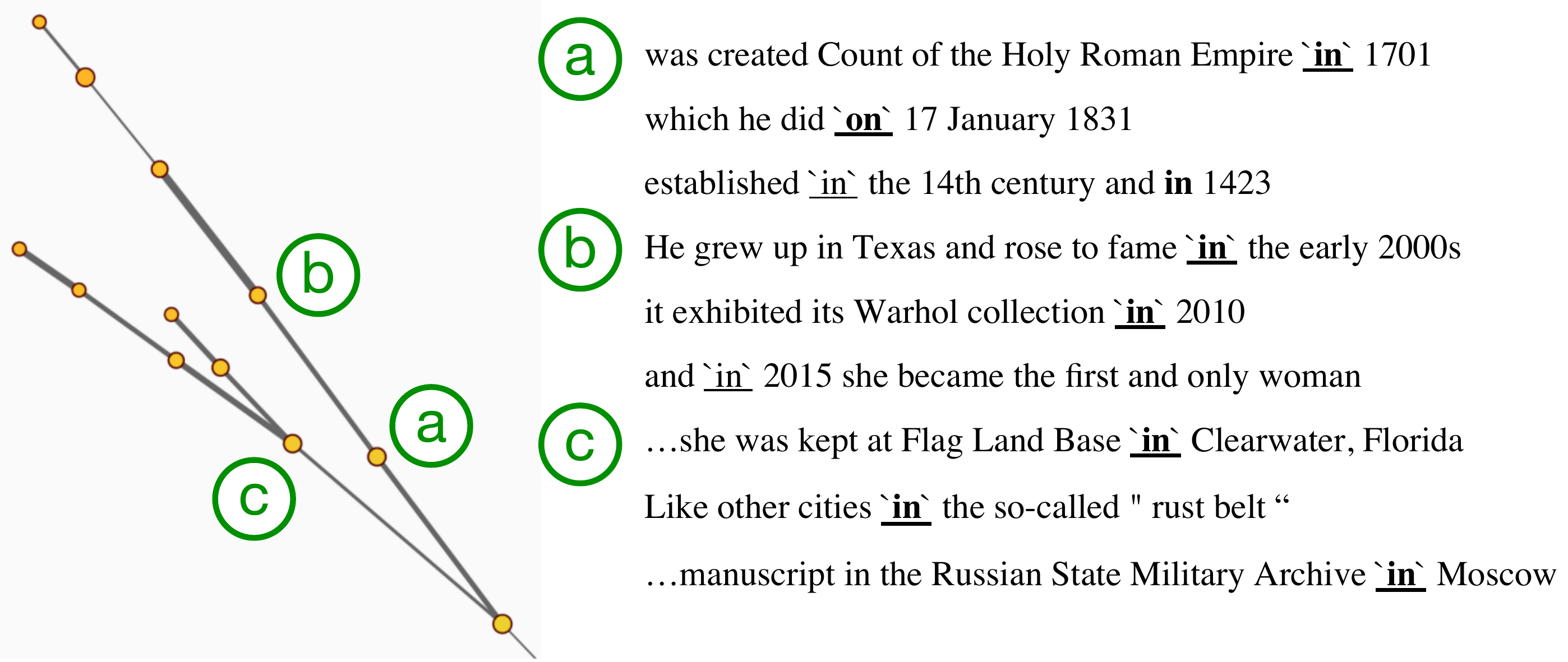}
}
\caption{Temporal and locative prepositions. Configuration: layer 9, Euclidean distance, 100 intervals, 30\% overlap.}
\label{fig:prepositions}
\vspace{-4mm}	
\end{figure}

The difference between temporal and locative usages of prepositions is well studied, and forms the basis of the Preposition Supersense Project that Dr. Srikumar is part of~\cite{SchneiderSrikumar2015}. 
As illustrated in~\autoref{fig:prepositions}, both nodes (a) and (b), and their parent node, represent clusters of prepositions, but the two branches capture distinct meanings rather than mere surface level differences. 
To see this, note that both branches include the preposition word \textbf{in}, but the branch (a) represents its temporal usage, whereas (b) represents its locative usage. 
This branching structure is reminiscent of the Supersense Hierarchy developed via linguistic analysis~\cite{SchneiderHwang2020a}. 

These, and the previous observations, suggest that we can discover linguistic structures from BERT activations using {\topoact}. 
Furthermore, {\topoact} also provides investigative directions for why these embeddings have been successful at a wide variety of linguistic tasks.

\section{Discussion}
\label{sec:discussion}

{\topoact} supports exploratory analysis of numerous interesting topological structures, locally and globally, in the space of activation vectors. 
We encourage readers to utilize the live demo of datasets for InceptionV1 and ResNet for such an exploration. 
Our approach is not without its limitations. 
The exploration scenarios presented here are specific to the choice of input images as well as the choice of activation vectors.
Further analysis is required to determine how stable the results are with respect to these choices. 
However, some of these limitations are common to other recent approaches (\emph{e.g.},\cite{CarterArmstrongSchubert2019,HohmanParkRobinson2020}). We offer some topics for discussion and future work.

\para{Generality.}
We focus on CNNs, specifically, InceptionV1 and ResNet-18, in this paper. 
However, our approach is not restricted to a particular network architecture. 
Mapper graphs could be generated and used as a vehicle for visual exploration whenever neuron activations are present.
{\topoact} can be generalized to explore new datasets coupled with other trained neural network architectures, such as ZFNet~\cite{ZeilerFergus2012}, AlexNet~\cite{KrizhevskySutskeverHinton2012}, and VGGNet~\cite{SimonyanZisserman2015}. 
We have showed that our approach is generalizable beyond image classifiers to include textual embedding networks such as BERT.

\para{Parameter tuning.} 
Practical and automatic parameter tuning for the mapper construction remains a challenging open problem for the broad TDA community. 
Carriere \etal~\cite{CarriereMichelOudot2018} provided the state-of-the-art, albeit theoretical, results on mapper parameter selection under restrictive settings.
Their framework assumed that a point cloud sample taken from the underlying space has a well-behaved, parameterizable probability distribution (formally, an $(a, b)$-standard distribution) and that the sample is sufficiently large, so that the Hausdorff distance between the sample and the underlying space is small.
However, upon careful investigation, these assumptions are not applicable in our setting.
Although we may assume that the activation space is a compact subset of the Euclidean space, we cannot verify that the activation vectors we sample follow the generative model of an $(a, b)$-standard distribution, and that $300K$ vectors form a sufficiently large sample for approximating or possibly reconstructing the underlying space.  

On the other hand, the mapper construction comes with ``best practices" in terms of parameter tuning, which rely on a grid search in the parameter space where good parameter combinations are those that produce stable structures.  Finding a theoretically sound and yet practical parameter tuning strategy for our mapper graph construction remains open; see the supplementary materials for more discussion on this topic. 
For the current version of the {\topoact}, we focus on exploring various mapper graphs with predetermined sets of parameter combinations following the best practices.
Additionally, confirmation bias is a risk when {\topoact} is utilized in practice because users may simply tune the parameters until they see what they want to see in the visualization. 
However, confirmation bias cannot be resolved without automatic parameter tuning, which remains an open problem.

\para{Stability.}
Additional theoretical results regarding the stability of mapper construction are available in~\cite{BrownBobrowskiMunch2019} and the references therein.
The investigation is on-going into how stable the mapper graphs are with respect to different sampling techniques. 
Under some assumptions on the sampling condition, Brown \etal~\cite{BrownBobrowskiMunch2019} showed that a pair of mapper graphs is close if their underlying point clouds are sampled from the same probability density function concentrated on the underlying topological space.
However, similar to the situation of parameter tuning, the gap between theory and practice is still large. 
Filling such a gap is beyond the scope of this paper. 

\para{Adversarial attacks.}
An important aspect in understanding the effectiveness of adversarial attacks on neural networks is how an attack alters the intermediate representations, \ie,~the activations. 
{\topoact} visualizes these representations from a topological perspective and hence might be useful in analyzing the effect of adversarial attacks at different layers of the  network.

\para{Corrective actions during training.}
A branching point (a bifurcation) in the space of activations at a particular layer may indicate the point where the network starts distinguishing a pair of classes. 
This knowledge can be useful to inform corrective actions for inputs in the test data that are being misclassified.
For example, if two classes that bifurcate at a particular layer in {\topoact} are still being misclassified as each other, an expert can choose to increase the network width at subsequent layers, or to selectively augment the training data for these classes to encourage better separation.

\section{Conclusion}
\label{sec:conclusion}

In this paper, we present {\topoact}, a framework to explore the topology of the activation spaces of neural networks.
We obtain topological summaries of the activation spaces via mapper graphs that capture the organizational principal behind neuron activations. 
We apply {\topoact} to trained neural networks such as ResNet and InceptionV1 for image classification, and BERT for contextual word embeddings. 
In each case, we present exploration scenarios that provide valuable insights into the image representations or word embeddings learned by different layers of these networks. 
This paper is the first step toward understanding the topological structure of the activation spaces in deep neural networks.


\section*{Acknowledgments}
This research was partially funded by NSF IIS 1910733, DBI 1661375, and IIS-1513616.  
We would like to thank Vivek Srikumar for his extremely valuable feedbacks regarding applying {\topoact} to the BERT neural networks. 
We also would like to thank Jeff Phillips who forwarded a Twitter message from Chris Olah on March 7, 2019, which inspired this project.  

\bibliographystyle{eg-alpha-doi}
\bibliography{Topo-ML.bib}

\clearpage
\begin{appendix}
	
	\section{{\topoact} User Interface and System Design}
\label{sec:interface}

We provide details regarding the user interface and system design of {\topoact}.
Figure 3 in the main paper illustrates the user interface under single-layer exploration mode. 

The control panel includes information regarding the layer of choice (e.g.,~\emph{3a}, \emph{3b}, \emph{4a}), the dataset (across various mapper parameters) under exploration (e.g., \emph{overlap-30-epsilon-fixed}, \emph{overlap-50-epsilon-adaptive}), and a class search box that supports filtering by a set of classes. 
It enables projections of the activation vectors using t-SNE and UMAP.  
The control panel also contains a check box that superimposes averaged activation images over the graph nodes to provide an alternative overview of the topological summary (see feature visualization panel for details). 
It also supports the filtering of graph edges based on the Jaccard index. 

\begin{figure}[!ht]
    \centering
    \includegraphics[width=.8\columnwidth]{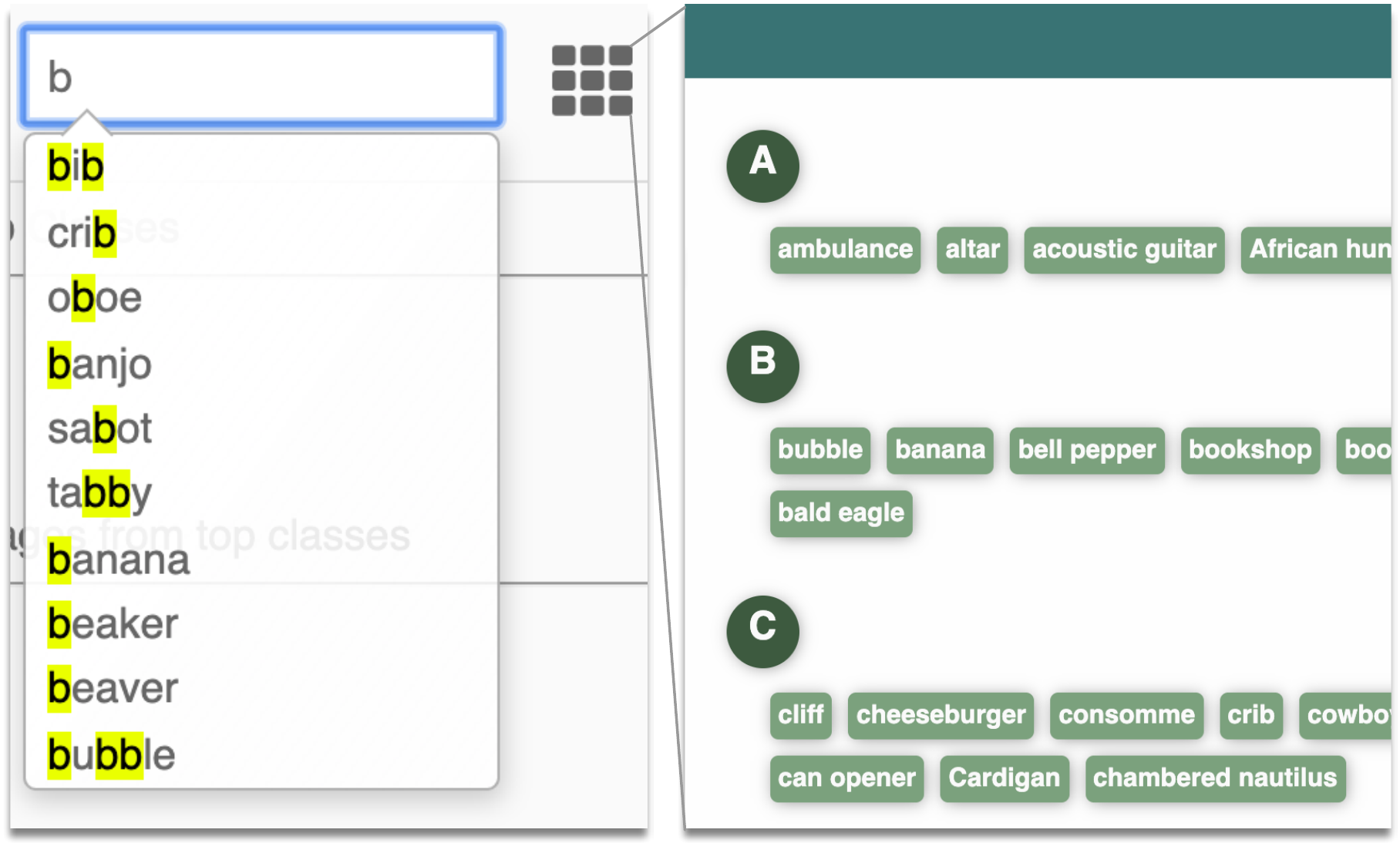}
    \caption{Class search box used to specify a set of classes to be filtered by the mapper graph.} 
\label{fig:class-search}
\end{figure}

\para{Class search box with a shopping directory view.}
As illustrated in~\autoref{fig:class-search}, users can type a class name in the search box, which is used to filter the mapper graph.
The search bar uses partial matching to locate a list of possible class names. 
Alternatively, users can select a subset of classes from the ``shopping directory'' view in which top classes within the current layer are listed in alphabetical order. 
The mapper graph will highlight the clusters that contain any of the user-specified classes among their top three classes.

\begin{figure}[!ht]
    \centering
    \includegraphics[width=.8\columnwidth]{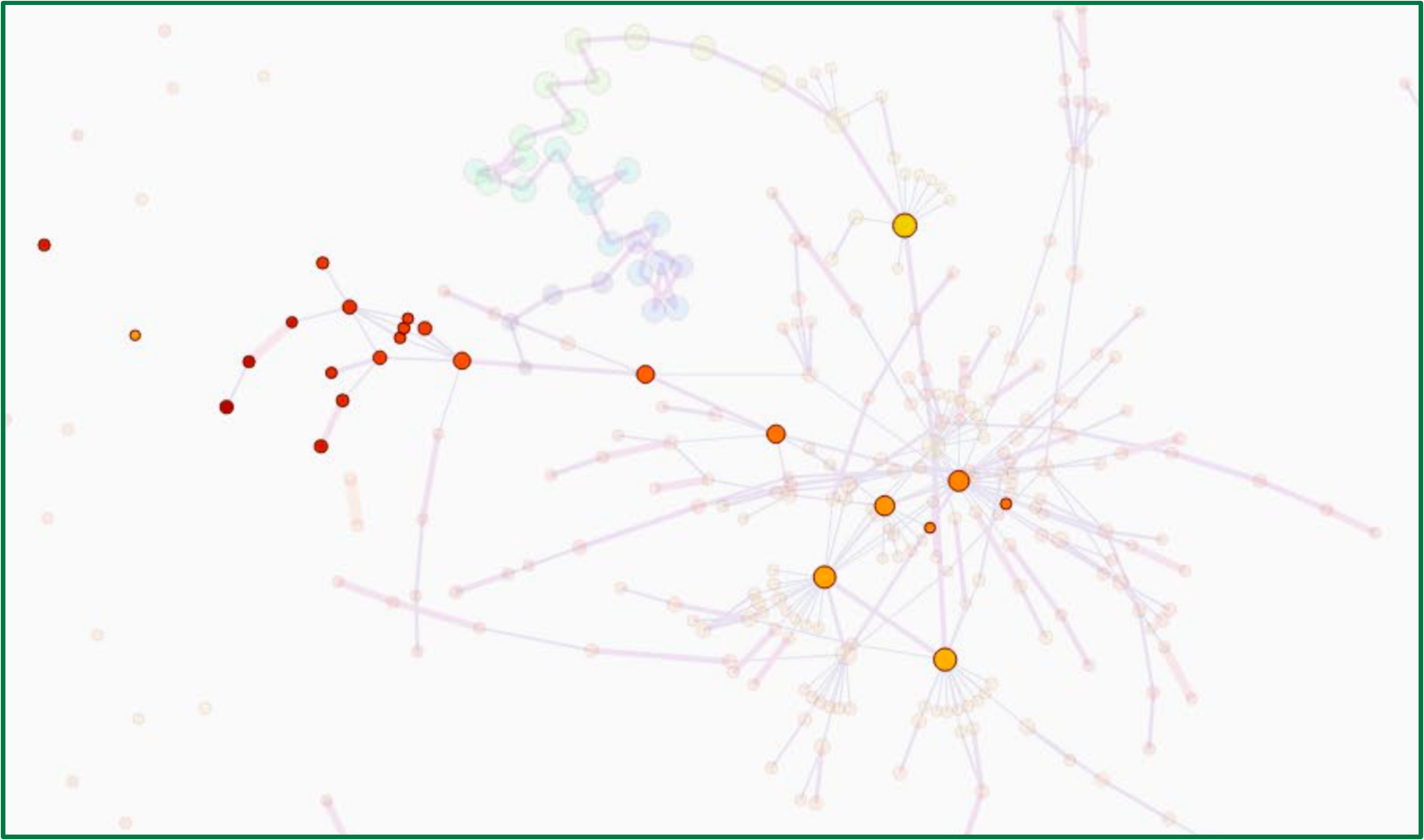}
    \caption{A mapper graph highlighting nodes that include classes of large motor vehicles.}
\label{fig:class-search-example}
\end{figure}

\begin{figure}[!ht]
	\centering
	\includegraphics[width=.6\columnwidth]{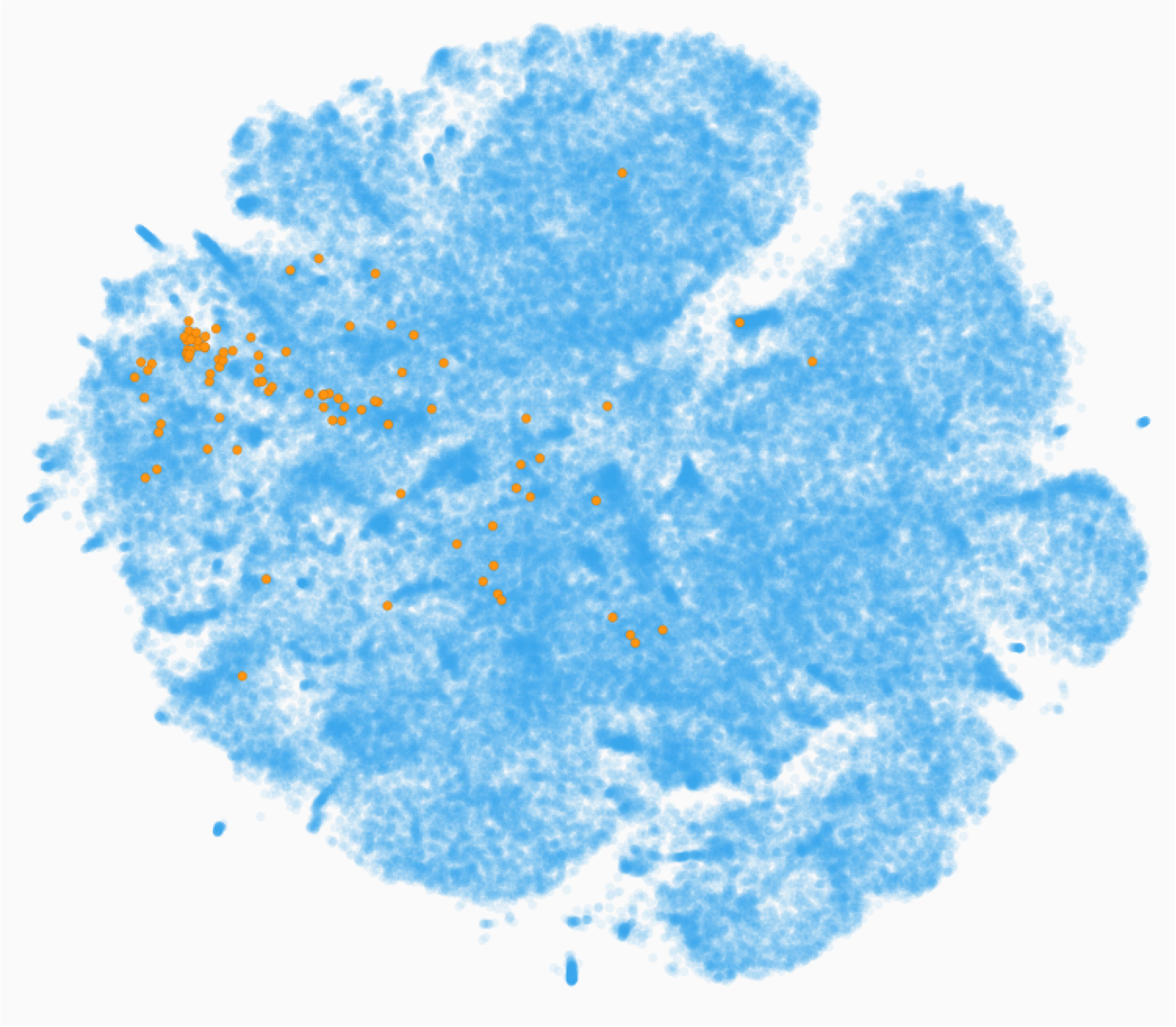}
	\vspace{-2mm}
	\caption{Class search in the projection view for the \textbf{lifeboat} class. Users can search one or more classes and visualize them in the t-SNE or UMAP projection.}
	\label{fig:tsne-class-search}
	\vspace{-4mm}
\end{figure}

When the projection view is enabled, class search will also highlight all activations for that class in the t-SNE/UMAP projection. 
As an example, in \autoref{fig:tsne-class-search}, we look at t-SNE projection of activations from layer \emph{5a} of the ImageNet dataset (\emph{overlap-30-epsilon-adaptive}).
Using the shopping directory view, we select several classes of large motor vehicles, for example, \textbf{school bus}, \textbf{tow truck}, \textbf{fire engine}, \textbf{minibus}, \textbf{minivan}, etc.
Each node highlighted in the mapper graph of \autoref{fig:class-search-example} contains at least one of the selected classes among its top three classes.

\subsection{Single Layer Exploration Mode}
For single layer exploration, the interface is composed of three panels: the mapper graph panel, the data example panel, and the feature visualization panel (see~Figure 3 in the main paper for an illustration). 

\para{Mapper graph panel.}
For ImageNet dataset, {\topoact} uses the mapper construction to construct a topological summary from the activation vectors of $300K$ images across $1K$ classes. 
Different from dimensionality reduction approaches such as t-SNE~\cite{MaatenHinton2008} and UMAP~\cite{McInnesHealyMelville2018}, {\topoact} computes and captures the shape of the activation space in the original high-dimensional space in the form of a mapper graph and preserves the structural information as much as possible when the mapper graph is drawn on the 2-dimensional plane. 

As shown in Figure 3(a) in the main paper, we use a force-directed layout by Dwyer~\cite{Dwyer2009} to visualize the mapper graph. 
Each node represents a cluster of ``similar'' activation vectors (in terms of their proximities in Euclidean distance), and each edge encodes the relations between clusters of activation vectors.
Given two clusters of activation vectors $C_u$ and $C_v$, an edge $uv$ connects them if $|C_u \cap C_v| \neq \emptyset$.
Given $C_u$ and $C_v$ connected by an edge $uv$, the edge weight of $uv$ is their Jaccard Index, that is, $J(C_u, C_v) := {|C_u \cap C_v|}/|C_u \cup C_v|$.
Each edge is then visualized by visual encodings (i.e.,~thickness and colormap) that scale proportionally with respect to their weights. 
Weights on the edges highlight the strength of relations between clusters. 

To explore the mapper graph, users can zoom and pan within the panel. 
Hovering over a node in the mapper graph will display simple statistics of the cluster: the number of activation vectors in the cluster and the averaged lens function value. 
Clicking on a node will give information on the top three classes (with a membership percentage) within the selected cluster; it will also update the selection for the data example panel and the feature visualization panel, as described below. 

\begin{figure}[!ht]
    \centering
    \includegraphics[width=.99\columnwidth]{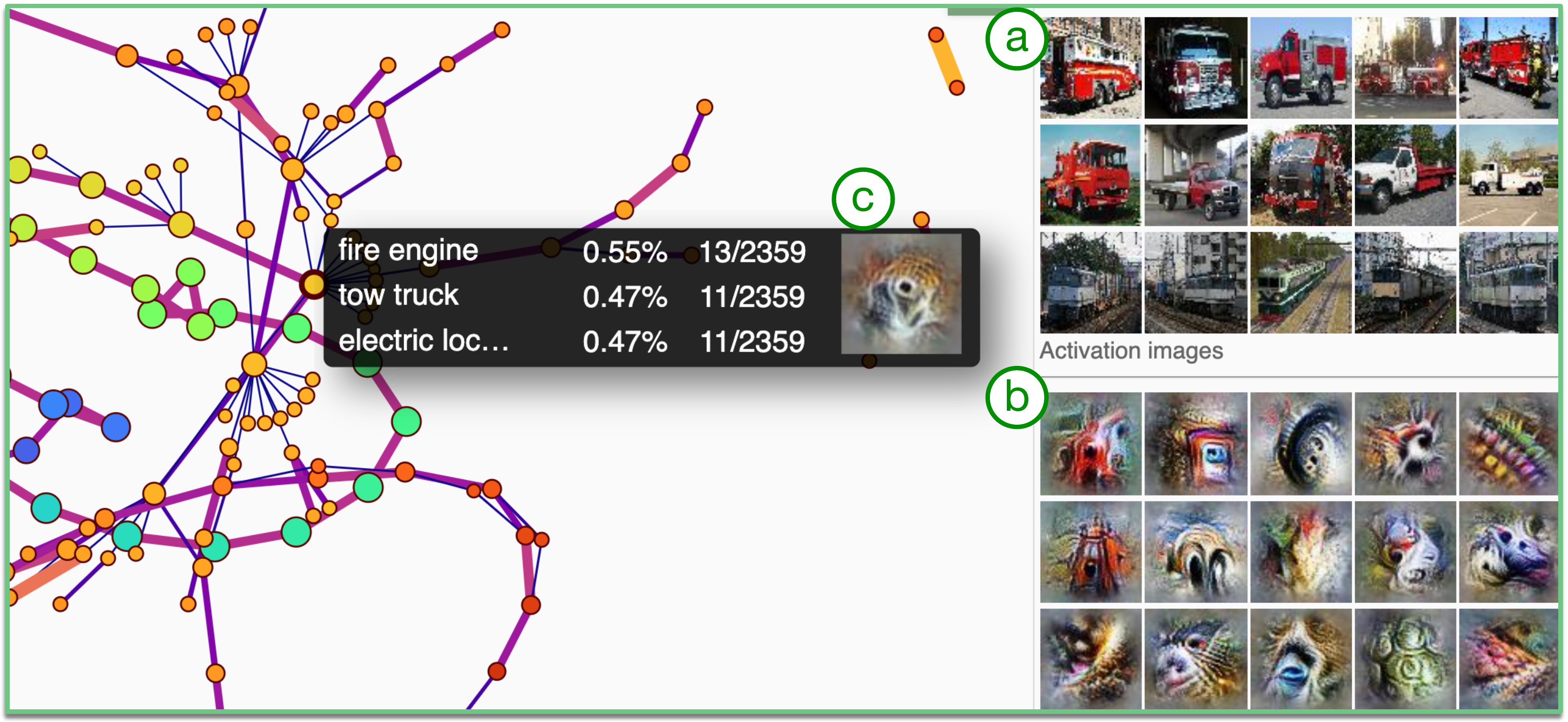}
    \caption{A data example panel (a) and a feature visualization panel (b) for layer 5a, where (c) contains an averaged activation image for the chosen cluster.} 
\label{fig:example-view}
\end{figure}

 \para{Data example panel.}
To make each cluster more interpretable, we combine the original data examples with feature visualization. 
For a selected node (cluster) in the mapper graph, we give a textual description of the top three classes in the cluster as well as five data examples from each of the three top classes. 
For example, as illustrated in~\autoref{fig:example-view}a, a selected cluster in the mapper graph view for layer \emph{5a} (of \emph{overlap-30-epsilon-adaptive})  contains the three top classes of images: \textbf{fire engine}, \textbf{tow truck}, and \textbf{electric locomotive}. 
Its corresponding data example view contains five images sampled from each class to give a concrete depiction of the input images that trigger the activations. 
 
\para{Feature visualization panel.}
After a user selects a node (cluster) in the mapper graph panel, we display activation images pre-generated for each input image from the data example panel. 
These individual activation images are generated by applying feature visualization to individual activation vectors from the $300K$ input images. 
The feature visualization displays up to $15$ of such individual activation images, up to $5$ for each of the top classes; see~\autoref{fig:example-view}(b). 
Furthermore, we also average the activation vectors that fall within the cluster and run feature inversion on the averaged activation, producing an \emph{averaged activation image} per cluster, as shown in~\autoref{fig:example-view}(c). 
Moving across clusters following edges of the mapper graph will help us understand how the averaged activation images vary across clusters. 
We obtain a global understanding of not only what the network ``sees'' via these idealized images but also how these idealized images are related to each other in the space of activations. 

In addition to the graph view, we can replace each node in the mapper graph by an averaged activation image as a glyph.
This can be perceived as an alternative to the \emph{activation atlas}~\cite{CarterArmstrongSchubert2019} with one crucial difference: the mapper graph captures clusters of activation vectors in their original high-dimensional space and preserves relations between these clusters.
Such a global view provides valuable insights during in-depth explorations. 

\para{t-SNE and UMAP projections.} 
For comparative purpose, we perform dimensionality reduction on the activation vectors for each layer using t-SNE and UMAP. 
The projection is done using all 300K activation vectors onto a 2-dimensional  space.
For t-SNE, we use the Multicore-TSNE~\cite{Ulyanov2016} Python library and set perplexity to be $50$ following the parameter choice used in the activation atlas~\cite{CarterArmstrongSchubert2019}. The UMAP projection is performed using its official Python implementation~\cite{McInnesHealySaul2018} with $20$ nearest neighbors and a minimum distance of $0.01$.
t-SNE and UMAP projections are precomputed due to the large number (300K) of activation vectors.  
We also provide a linked view between the mapper graph and the t-SNE/UMAP projection. Selecting a node in the mapper graph will highlight its corresponding activation vectors in the t-SNE/UMAP projections. 
We provide subsampled versions of these projections (5K, 10k, 50K, 100K, and 300K) to deal with the issue of visual clutter and to accommodate various browser rendering capabilities on a number of devices.

\subsection{Multilayer Exploration Mode}
\begin{figure}[!ht]
    \centering
    \includegraphics[width=.98\columnwidth]{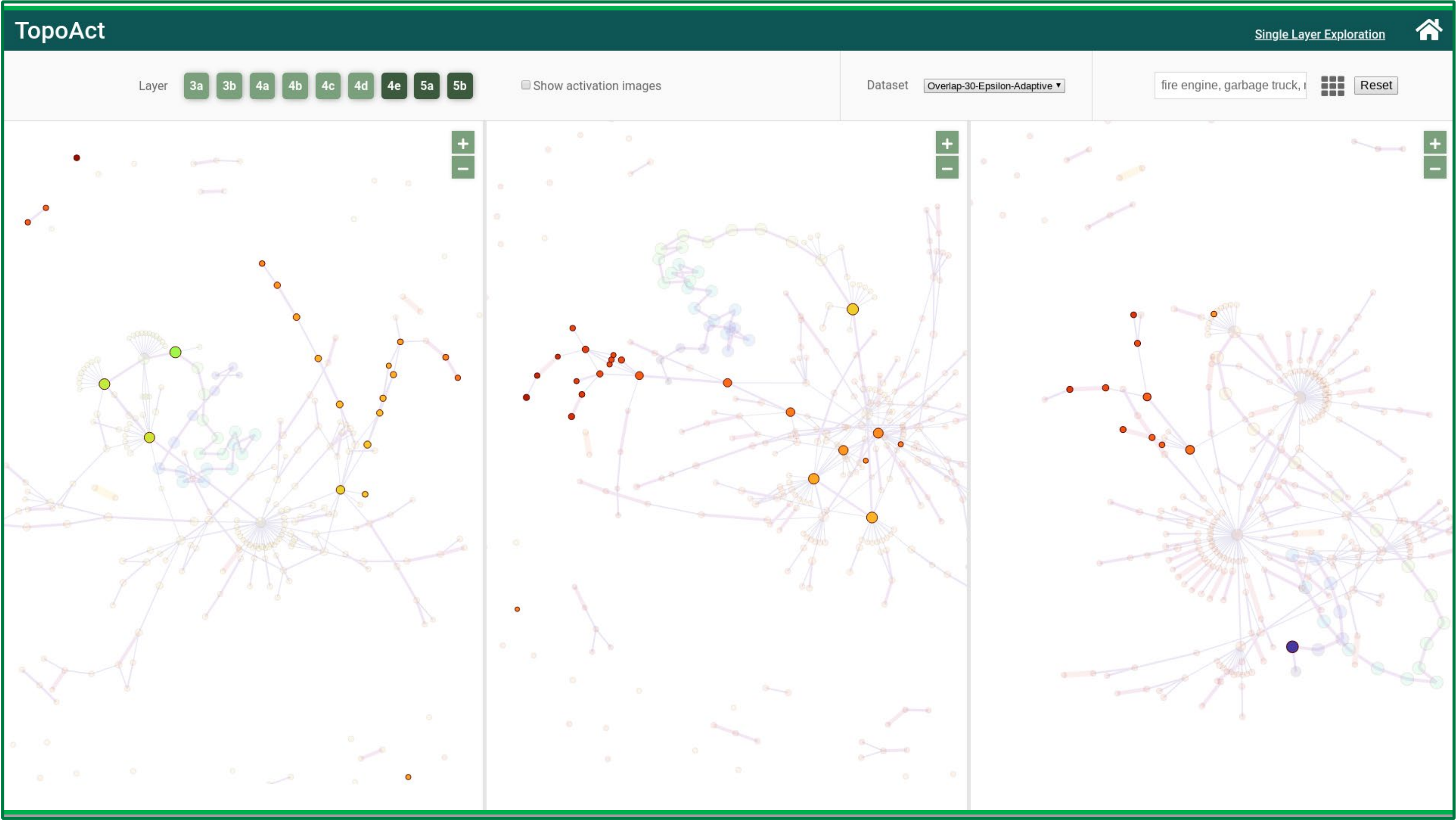}
    \caption{Class search highlights nodes that include classes of large motor vehicles across multiple layers.}
\label{fig:class-search-multilayer}
\vspace{-4mm}
\end{figure}

\begin{figure}[!ht]
    \centering
    \includegraphics[width=.99\columnwidth]{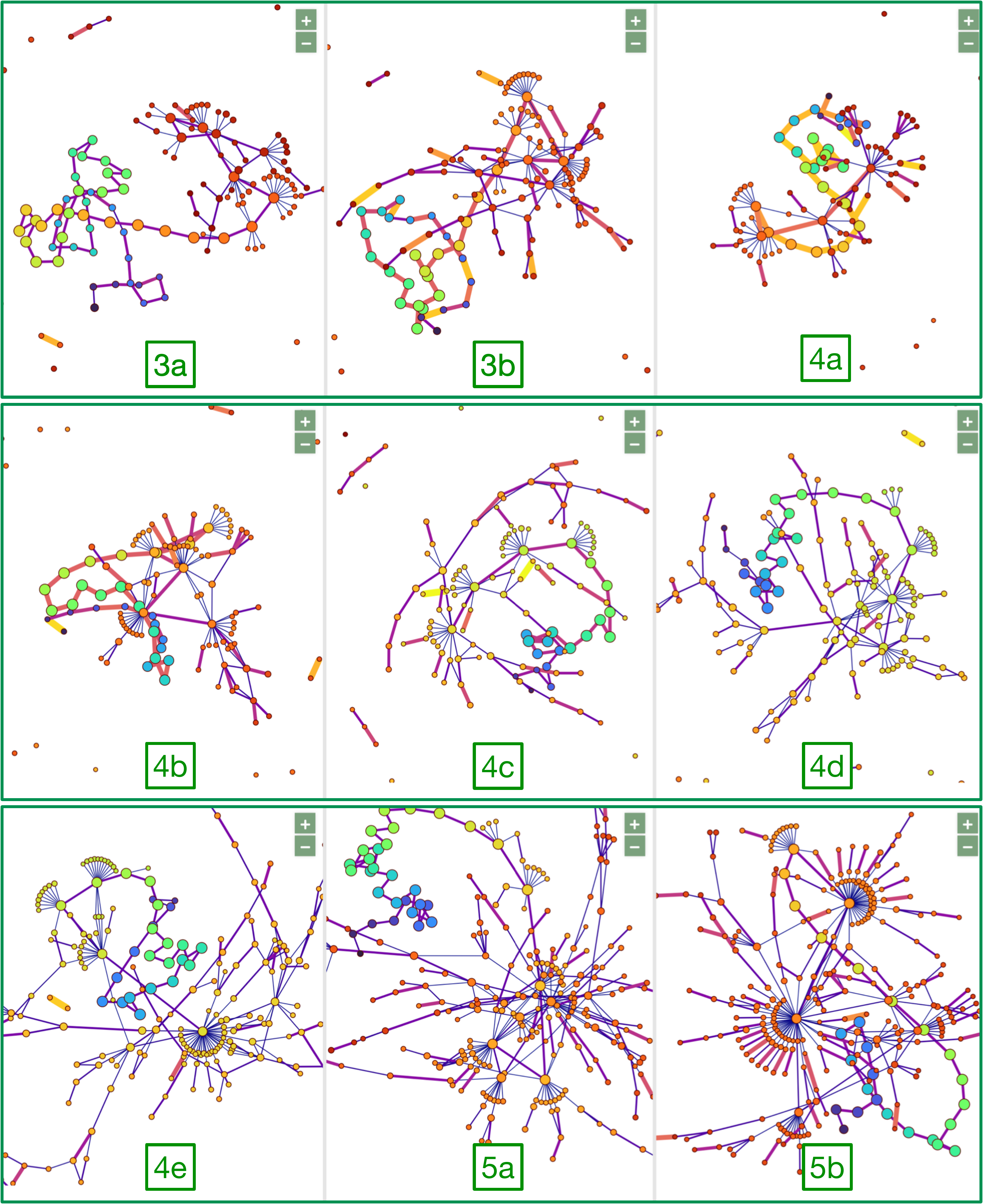}
    \caption{Comparing nine mapper graphs for the ImageNet dataset using multilayer exploration. Configuration: $70$ intervals, $30\%$ overlap, adaptive $\epsilon$ for DBSCAN.} 
\label{fig:global-all-layers}
 \vspace{-4mm}
\end{figure}

In the multilayer exploration mode, three adjacent layers are explored side by side; see~\autoref{fig:global-all-layers}(top). 
After choosing a particular class or a set of classes using the class search box, {\topoact} highlights nodes (clusters) across all three layers that contain the chosen set of classes among its top three classes. 
Other visualization features are inherited from the single layer exploration. 
Multilayer exploration helps capture the evolution of classes as images are run through the network and supports structural comparisons of summaries across layers.
Such exploration can be particularly useful when used in conjunction with the class search tool. 
As an example of a class search in multilayer mode, we look at layers \emph{4e}, \emph{5a} and \emph{5b} of the \emph{overlap-30-epsilon-adaptive} dataset.
We use the same selection of classes of large motor vehicles used in the earlier example of a class search in single layer mode (\autoref{fig:class-search-example}).
\autoref{fig:class-search-multilayer} shows the class search results, now in the multilayer exploration mode.

Under the multilayer exploration mode, we can compare the shape of activation spaces across multiple layers. 
As illustrated in~\autoref{fig:global-all-layers}, we show a side-by-side comparison of all layers for the ImageNet dataset (\emph{overlap-30-epsilon-adaptive}).
A further investigation into structural comparisons across layers, such as tracking the evolution of a particular branching node, is nontrivial and left for future work.

\subsection{System Design}
{\topoact} is open-source via GitHub: \url{https://github.com/tdavislab/TopoAct/}, and web-based with a public demo: \url{https://tdavislab.github.io/TopoAct/}. 
It is tested for Google Chrome and Mozilla Firefox. 
It is developed using Javascript, HTML, and CSS, along with D3.js and Chart.js. 
The $300K$ dataset examples were sampled from ImageNet dataset.  
For our mapper graph construction, we used a modified version~\cite{ZhouChalapathiRathore2020} of the open-source KeplerMapper library~\cite{VeenSaul2019} that we optimized to handle the large number of data points that we encountered in our use case.
The construction of mapper graphs across layers was performed on high-performance server machines with $128$, $160$, and $256$ CPU cores, and RAM ranging from $504$ GB to $1024$ GB. 
The construction took around 15 minutes for layers with lower dimensional activation vectors (i.e.,~layer \emph{3a} produces 256-dimensional activation vectors) and 25-30 minutes for higher dimensional activation vectors (e.g.,~layer \emph{5b} produces 1024-dimensional  activation vectors).
For our choice of $\epsilon$ for the DBSCAN algorithm, we ran   PyNNDescent~\cite{PyNNDescent} on a commodity workstation with a 4 core intel i7 (4750HQ) and 8GB of RAM.
Computing $\epsilon$ took on average 5 minutes per layer. 
Finally, we used Google Colab~\cite{Bisong2019, GoogleColab} to run our feature visualization with GPUs, either from an Nvidia P100, Nvidia K80, or Nvidia T4 GPU.  
Feature visualization of all 300K input images was done via the Lucid library~\cite{Lucid}, which took on average 8 hours. 
Feature visualization of average activation vectors took between 2.5 (i.e.,~\emph{3a}) and 6 hours (i.e.,~\emph{5b}) per mapper graph.

	\section{$L_2$ Norm and Adaptive Cover}
\label{sec:l2norm}

In the demo, we used a uniform cover, which caused large variations in cluster sizes.
Although some clusters were composed of only a handful of activation vectors, several clusters had thousands of activation vectors, and large intersections between neighboring clusters.
Finding meaningful relationships across such large clusters is difficult in these cases since the top three classes may not be good representatives of the cluster as a whole.

The branches and loops explored in our examples contain relatively small clusters for which the averaged activation images are more meaningful.
The best way to remedy the large variation in cluster sizes is to use an adaptive cover, in which interval lengths are modified in such a way that each interval contains approximately the same number of points. 
Creating adaptive cover elements may be achieved by looking at the distribution of lens function values using histograms. 
We now discuss this in more detail.

In general, vectors with a dimension as high as the ones from a neural network (maximum of 1024 dimensions in our case) tend to suffer from the curse of dimensionality, which implies that in very high dimensions, the Euclidean metric or the $L_2$ norm does not exhibit variation - all distances and norms look the same.

\begin{figure}[!h]
    \centering
    \includegraphics[width=0.95\columnwidth]{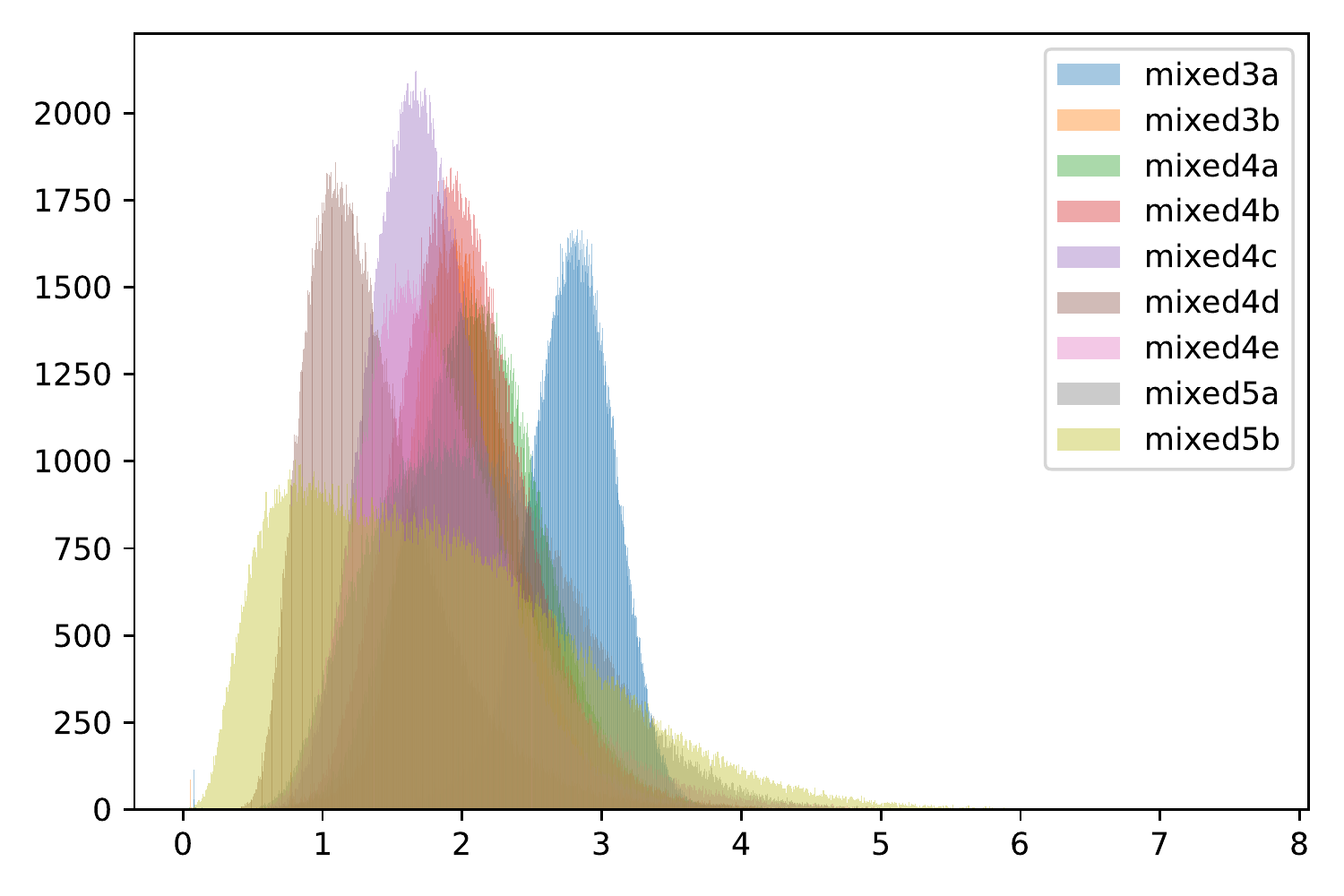}
    \caption{$L_2$ norms of activation vectors across all layers.} 
\label{fig:overlayed_l2norms}
\end{figure}

\autoref{fig:overlayed_l2norms} shows the distribution of $L_2$ norms for all layers in the Inception architecture. 
Notice that the distribution is bell-shaped with long tails, and the variance of the distribution is reasonably large. 
The severity of the curse of dimensionality may be reduced by using an adaptive cover that has more intervals in the denser regions of lens function. 
The resulting mapper graphs with such an adaptive cover will contain nodes of comparable sizes. 
This is left for future work.

\end{appendix}


\end{document}